\documentclass[preprint]{aastex}

\slugcomment{Submitted to the Astrophysical Journal}

\shorttitle{Cardall, Prakash, \& Lattimer}
\shortauthors{Effects of Strong Magnetic Fields on Neutron Star
	Structure}

\begin{document}

\title{Effects of Strong Magnetic Fields on Neutron Star
	Structure}

\author{Christian Y. Cardall,
	Madappa Prakash, and James M. Lattimer}
\affil{Department of Physics \& Astronomy, State University of New
	York at Stony Brook, Stony Brook, NY 11794-3800}

\date{November 6, 2000}

\begin{abstract}
We study static neutron stars with poloidal magnetic fields
and a simple class of electric current distributions consistent 
with the requirement of stationarity.  
For 
this
class of electric current 
distributions, we 
find that magnetic fields are too large for static configurations to
exist when the magnetic force pushes a sufficient amount of mass
off-center that the gravitational force points outward near the origin
in the equatorial plane.  (In our coordinates an outward gravitational
force corresponds to $\partial\ln g_{tt}/\partial r>0$, where $t$ and
$r$ are respectively time and radial coordinates and $g_{tt}$ is
coefficient of $dt^2$ in the line element.)  For the equations of
state (EOSs) employed in previous work, we obtain configurations of
higher mass than had been reported; we also present results with more
recent EOSs.  For all EOSs studied, we find that the maximum mass among
these static configurations with magnetic fields is
noticeably larger than the maximum mass attainable by uniform
rotation, and that for fixed values of baryon number
the maximum mass configurations are all characterized by an off-center
density maximum.
\end{abstract}


\section{INTRODUCTION}
\label{sec:intro}

Over the years, the typical magnitudes of the surface magnetic fields of
pulsars---as inferred from measured spindown rates and simple magnetic dipole
models---have been $\sim 10^{12}-10^{13}$ G 
\citep{tayl93}.  
Assuming magnetic flux conservation, fields of $\sim 10^{12}$ G would 
naturally arise from 
typical main sequence star
surface magnetic fields of $\sim 100$ G during a
decrease in radius by a factor of $\sim 10^5$ \citep{shap83}.
At the extreme end of fields attainable by flux conservation, 
the largerst observed white dwarf magnetic 
field of $5\times 10^8$ G leads to a neutron star field of 
$\sim 1\times 10^{14}$ G \citep{carr96}, while the largest
observed main sequence stellar magnetic field of $3.4 \times 10^4$~G 
\citep{bohm89} also suggests a
possible field of a few times $10^{14}$~G. 

Several independent
circumstantial arguments link the class of objects known
as ``soft $\gamma$-ray repeaters'' (SGRs), and perhaps 
the so-called ``anomalous X-ray pulsars'' (AXPs),
with neutron stars having magnetic fields 
$\gtrsim 10^{14}$ G---the so-called ``magnetars''
\citep{dunc92,usov92,pacz92,thom95,thom96,vasi97a}.  
(Table \ref{tbl-0} displays some observed properties of these objects.)
In addition to the circumstantial arguments,
more direct evidence for magnetic fields of $2-8\times 10^{14}$ G
is available for two of the 
five
known SGRs,  
in the form of measured periods and
spindown rates of associated X-ray pulsars  
\citep{kouv98,kouv99}.\footnote{For ongoing discussion
of this interpretation of X-ray timing data see e.g.
\cite{mars99a,wood99a,hard99,mars99b,chat00}.}
Furthermore, the observed X-ray luminosities of 
the AXPs
may require a field strength $B\gtrsim 10^{16}$ G
\citep{chat00,heyl98b}.
The population statistics of SGRs suggest that magnetars may 
constitute a significant fraction ($\gtrsim 10\%$) of the neutron star
population \citep{kouv94,kouv98}. 
As mentioned above, there
are isolated examples of progenitor stars which could yield 
fields of $\sim 10^{14}$ G by flux conservation, but these isolated
examples do not seem sufficient to account for a significant fraction
of the neutron star population. Thus, it seems likely that some
mechanism {\em generates} magnetic fields in nascent neutron stars.
For example, \cite{dunc92} suggest that the smoothing out
of differential
rotation and convection could generate fields as large as $3\times 10^{17} 
(P_i/1\ \rm{ms})^{-1}$ G, where $P_i$ is the initial rotation period of
the neutron star.

These considerations motivate study of the effects of ultra-strong
magnetic fields on neutron star properties. 
In this we have been inspired by the pioneering work of
Bocquet, Bonazzola, Gourgoulhon \& Novak (1995), who performed
relativistic calculations of axisymmetric neutron star structure in
which the standard stress-energy tensors of a perfect fluid and the
electromagnetic field were employed, and were comparable in magnitude.
The maximum fields they found were of order $10^{18}$ G, with
increases of 13 to 29\% in the maximum mass of nonrotating stars for
various equations of state.

An additional motivation is provided by the recent findings that
magnetic fields of strengths larger than $10^{16}$ G affect the EOS of
dense matter directly through drastic changes in the composition of
matter \citep{cha96,cha97,yz99,bpl00}.  The EOS is altered by both the
Landau quantization of the charged particles (such as
protons, electrons, etc.) and the interactions of the magnetic
moments, including the anomalous magentic moments of the neutral
particles (such as the neutron, strangeness-bearing $\Lambda$-hyperon
etc.) with the magnetic field.  In this work, we consider
only the effects of the magnetic field on the
structure, through its influence on the metric, in
order to facilitate a comparison with the earlier work of
\cite{bocq95}.  The additional changes caused by the direct effects of
the magnetic field on the EOS will be reported in a future work
\citep{card00}.

While \cite{bocq95} also presented solutions for rotating neutron
stars endowed with large magnetic fields, in this work we present only
static solutions. In terms of the potential observability of the
effects of large magnetic fields, the most relevant situation appears
to be for non-rotating stars.  Neutron stars with the highest
inferred magnetic fields---the so-called ``magnetars''---are all
observed to be rotating very slowly, with periods on the order of
seconds.  The effects of such slow rotation should have a negligible
impact on the neutron star structure.

In this paper we extend the work of \cite{bocq95}.  The
theoretical formalism is outlined in \S\ref{sec:form}, which serves to
put the problem in context.  In \S\ref{sec:himass} we shed light on an
issue left somewhat unclear by \cite{bocq95}: What physically
determines the maximum mass and magnetic field for a given maximum
density or given baryon mass?  In order to explore these questions we
have chosen to solve the structural equations using a Green's function
technique rather than the spectral technique employed by
\cite{bocq95}, and we also searched for the maximum mass, for a given
magnetic field distribution, in a different way.  An appendix
describes our numerical methods and tests of our code.  In
\S\ref{sec:constmb} we present an illuminating view of constant baryon
mass and constant magnetic moment sequences, and present higher mass
configurations than those found by \cite{bocq95} for the equations of
state (EOSs) they employed.  In addition, we present the results of
analogous calculations using three more recent EOSs.  Summary and
outlook are contained in \S\ref{sec:sumout}.

\section{FORMALISM}
\label{sec:form}

We consider stationary neutron star models in
which the equation of state is independent of the magnetic
field. The relevant equations and some properties of
neutron stars in this limit were studied by
\cite{bona93} and \cite{bocq95}.
The stress-energy tensor is given by the sum of
the standard stress-energy tensors of a perfect
fluid and the electromagnetic field: 
\begin{equation}
T^{\alpha\beta}=\left(e + p\right) u^\alpha u^\beta  + p\, g^{\alpha\beta} 
 + {1\over 4\pi}
	\left(F^{\alpha\rho} F^\beta_{\; \; \rho} - {1\over 4} g^{\alpha\beta}
	F_{\rho\sigma} F^{\rho\sigma}\right), \label{seunmag}
\end{equation}
where $e$ and $p$ are respectively the rest-frame energy density
and pressure, $u^\alpha$ is the fluid 4-velocity, $g_{\alpha\beta}$ are
the metric components,
and 
$F_{\alpha\beta}\equiv A_{\beta,\alpha}-A_{\alpha,\beta}$, where
$A_\alpha$ is the electromagnetic potential 1-form. 

At least two relativistic 
formulations of the problem of strongly nonspherical axisymmetric
stars have been developed.
Most authors
studying rapid rotation have adopted the approach 
of \cite{bard71}, which
explicitly assumes an isotropic stress
tensor and is thus incompatible with
electromagnetic fields. Building on earlier work
of \cite{bona71} and \cite{bona74}, \cite{bona93}
present a formulation which allows for the 
most general stress-energy tensor consistent
with a spacetime having the properties of 
stationarity, axisymmetry, and circularity.
The metric for such a spacetime can be expressed
as
\begin{equation}
g_{\alpha\beta}dx^\alpha dx^\beta = -e^{2\nu}dt^2 + e^{-2\nu}G^2
	r^2 \sin^2 \theta (d\phi - N^\phi dt)^2
	+ e^{2(\zeta-\nu)}(dr^2 + r^2 d\theta^2),\label{metric}
\end{equation} 
where the metric functions $\nu$, $G$, $N^\phi$,
and $\zeta$ are functions of $(r,\theta)$.
(A spacetime having the properties of stationarity and
axisymmetry, but not circularity,
would have an additional off-diagonal term in
the metric \citep{bocq95,cart73}).
For the spacetime to have the property of circularity 
in addition to the properties of stationarity 
and axisymmetry, 
it is necessary that the electromagnetic current
4-vector and fluid 4-velocity be parallel to a general
linear combination of the Killing
vectors \citep{cart73}. For example, 
in the coordinates of equation (\ref{metric}),
$J^t$ and $J^\phi$ would be the only nonvanishing 
components of the electromagnetic
current 4-vector. A further consequence of this is that
$A_t$ and $A_\phi$ are the only nonvanishing components
of the electromagnetic potential 1-form \citep{cart73}.

From the Einstein equations, \cite{bona93} 
derive a Poisson equation for each 
of the metric variables, with source terms
that depend on the metric variables and on 
the components of the stress-energy tensor:
{\samepage
\begin{eqnarray}
\Delta_3\, \nu &=& \sigma_\nu, \label{d1}\\
\tilde\Delta_3\, \tilde N^\phi &=&\sigma_{\tilde N^\phi}, \\
\Delta_2\, \tilde G &=& \sigma_{\tilde G}, \label{d3} \\
\Delta_2\, \zeta &=& \sigma_\zeta, \label{d4}
\end{eqnarray}
}
where
{\samepage
\begin{eqnarray}
\tilde N^\phi &\equiv & r \sin\theta\, N^\phi,\\
\tilde G &\equiv & r \sin\theta\, G,
\end{eqnarray}
}
and $\Delta_2$, $\Delta_3$, and $\tilde\Delta_3$ are
respectively the 2D flat space Laplacian, the 3D flat space
Laplacian, and the $\phi$ component of the 3D flat space
vector 
Laplacian:
\begin{eqnarray}
\Delta_2 &\equiv& {\partial^2\over \partial r^2} + {1\over r}
	{\partial\over \partial r} + {1\over r^2}{\partial^2 \over
	\partial \theta^2}, \\
 \Delta_3 &\equiv& {\partial^2\over \partial r^2} + {2\over r}
	{\partial\over \partial r} + {1\over r^2}{\partial^2 \over
	\partial \theta^2} + {1\over r^2 \tan\theta}{\partial\over
	\partial\theta}, \\
\tilde \Delta_3 &\equiv& \Delta_3 - {1\over r^2 \sin^2 \theta}.
\end{eqnarray}
The source terms in equations (\ref{d1}-\ref{d4}) 
involve the metric functions and components of
the stress-energy tensor:
\begin{eqnarray}
\sigma_\nu &=& 4\pi G_N \,e^{2(\zeta-\nu)}(E + S^i_{\ i})
	 + {1\over 2}\, e^{-4\nu}
	G^2 r^2 \sin^2 \theta \,(\partial N^\phi)^2 
 	  -\partial\nu\,\partial(\ln G), \label{snu} \\
\sigma_{\tilde N^\phi}&=&-{16\pi G_N\, e^{2\zeta+\nu} \over G^2}
	{I_\phi \over r\sin\theta}- r\sin\theta\, \partial N^\phi\,
	\partial\left[\ln\left(e^{-4\nu}G^3\right)\right], \\
\sigma_{\tilde G}&=& 8\pi G_N\, e^{2(\zeta-\nu)} G r\sin\theta   
	(S^r_{\ r} + S^\theta_{\ \theta}), \label{sg} \\
\sigma_\zeta &=& 8\pi G_N\, e^{2(\zeta-\nu)} S^\phi_{\ \phi}
	 +{3\over 4} e^{-4\nu}
	G^2 r^2 \sin^2 \theta (\partial N^\phi)^2  -(\partial\nu)^2.
	\label{szeta}
\end{eqnarray} 
In these expressions the notation $\partial X\, \partial Y$ denotes
\begin{equation}
\partial X\, \partial Y\equiv {\partial X \over
	\partial r}{\partial Y\over \partial r} + 
	{1\over r^2}{\partial X \over \partial \theta}
	{\partial Y \over \partial \theta}.
\end{equation}
In addition, $G_N$ is Newton's constant, and the
contributions from the stress-energy tensor are\footnote{We 
do not use $E$ to denote the magnitude
of the electric field, although we will use the subscripted
notation $E_i$ to refer
to components of the electric field.}
\begin{eqnarray}
E&=&T_{\alpha\beta}n^\alpha n^\beta, \\
I_\alpha &=& -h_{\alpha\beta} n_\rho T^{\beta\rho},\\
S_{\alpha\beta}&=& h_{\alpha\rho} h_{\beta\sigma} T^{\rho\sigma}, 
\end{eqnarray}
where $n^\alpha$ is the unit four vector orthogonal to the
spacelike time slices and 
$h_{\alpha\beta}=g_{\alpha\beta} +n_\alpha n_\beta$ is
the spacelike projection tensor.
In the present coordinates
$n_\alpha=(-N,0,0,0)$, where $N=e^\nu$ is the lapse function.
For the stress-energy tensor of a perfect fluid
[the first part of equation (\ref{seunmag})],
\begin{eqnarray}
E^{\rm PF}& =& \Gamma^2(e+p)-p, \\
\left(I^{\rm PF}\right)_\phi &=& e^{-\nu} G\, r\sin\theta\, 
	(E^{\rm PF}+p)\, U, \\
\left(S^{\rm PF}\right)^r_{\ r} &=& p, \ \ \ 
	\left(S^{\rm PF}\right)^\theta_{\ \theta}
	=p, \ \ \ 
	\left(S^{\rm PF}\right)^\phi_{\ \phi}=p+(E^{\rm PF}+p)\,U^2,
\end{eqnarray}
where $\Gamma = (1-U^2)^{-1/2}$ and $U=e^{-2\nu}G r\sin\theta
(\Omega - N^\phi)$, 
and the superscript ``PF'' stands for ``perfect fluid.''
In this expression for $U$, the scalar $\Omega$ is the
angular velocity as measured by a static observer at infinity;
it relates the nonvanishing components of $u^\alpha$
through the equation $u^\phi=\Omega\, 
u^t$.\footnote{As will be discussed below, in the limit of 
infinite conductivity (``frozen-in magnetic fields''), 
stationary stars with magnetic fields
must be uniformly rotating.}
For the standard stress-energy tensor of the electromagnetic
field [the last two terms of equation (\ref{seunmag})],
restricted to a poloidal field, we have
\begin{eqnarray}
E^{\rm EM}& =& {1\over 8\pi}(E_i E^i + B_i B^i), \\
\left(I^{\rm EM}\right)_\phi &=& 
	{1\over 4\pi}\,e^{2\zeta-3\nu} G r^2 \sin\theta
	(E^r B^\theta-E^\theta B^r), \\
\left(S^{\rm EM}\right)^r_{\ r} &=& {1\over 8\pi}(E_\theta E^\theta
	-E_r E^r + B_\theta B^\theta - B_r B^r),\\  
	\left(S^{\rm EM}\right)^\theta_{\ \theta}
	&=& - \left(S^{\rm EM}\right)^r_{\ r}, \ \ 
	\left(S^{\rm EM}\right)^\phi_{\ \phi}=E^{\rm EM},
\end{eqnarray}
in which the superscript ``EM'' identifies the electromagnetic contributions.
In these expressions $E_i$ and $B_i$ are the components of
the electric and magnetic fields as measured by an Eulerian
observer (i.e. an observer with four velocity 
$n^\alpha$):
\begin{eqnarray}
E_\alpha &=& n^\beta F_{\alpha\beta} \\
 &= & \left(0,e^{-\nu} \left[{\partial A_t \over \partial r} +
	N^\phi {\partial A_\phi\over \partial r}\right],
	e^{-\nu}\left[{\partial A_t\over \partial\theta} +
	N^\phi {\partial A_\phi \over \partial\theta}\right],0\right), \\
B_\alpha &=& -{1\over 2}\,\epsilon_{\alpha\beta\rho\sigma} n^\beta 
	F^{\rho\sigma} \\
	&=& \left(0,{e^\nu \over G r^2 \sin\theta}{\partial A_\phi
	\over \partial \theta}, -{e^\nu \over G\sin\theta}
	{\partial A_\phi\over \partial r},0\right),
\end{eqnarray}
where $\epsilon_{\alpha\beta\rho\sigma}$ is the Levi-Civita tensor.

The quantities $A_t$ and $A_\phi$
are also determined by Poisson equations, which derive from the Maxwell
equations in curved spacetime \citep{bocq95}:
\begin{eqnarray}
\Delta_3 A_t &=& \sigma_{A_t}, \label{ateq}\\
\tilde \Delta_3 \tilde A^\phi &=& \sigma_{\tilde A^\phi},\label{apheq}
\end{eqnarray}
where $\tilde A^\phi \equiv A_\phi / (r\,\sin\theta)$. 
The sources can be expressed
\begin{eqnarray}
 \sigma_{A_t}& =& -4\pi\; e^{2(\zeta-\nu)}\left(g_{tt}J^t + g_{t\phi}J^\phi
	\right) + e^{-2\nu}g_{t\phi} \,\partial A_t\, \partial N^\phi
	- \left(2+e^{-2\nu} g_{tt}\right)\partial A_\phi \partial N^\phi
	\nonumber \\
	& &-(\partial A_t + 2N^\phi \partial A_\phi)\,\partial\left[\ln
	\left(e^{-2\nu} G \right)\right] - {2 N^\phi\over r}
	\left({\partial A_\phi\over \partial r} + {1\over r \tan\theta}
	{\partial A_\phi \over \partial \theta}\right), \label{sat}\\
\sigma_{\tilde A^\phi}&=&-4\pi\; e^{2\zeta-4\nu} G^2 r\sin\theta
	\left(J^\phi - N^\phi J^t\right) + e^{-4\nu} G^2 r\sin\theta\,
	\partial N^\phi \left(\partial A_t + N^\phi \partial A_\phi\right)
	\nonumber\\
	& &+{1\over r\sin\theta}\partial A_\phi \partial\left[\ln
	\left(e^{-2\nu} G \right)\right]\label{saph}.
\end{eqnarray}

A theorem of \cite{cowl34} states that an axisymmetric magnetic
field cannot be generated or maintained by the motion of a fluid.
The theorem relies on the fact that
finite resistivity involves dissipation, 
leading to magnetic field decay.\footnote{Ambipolar diffusion and 
Hall drift \citep{gold92} will actually dominate
magnetic field decay in magnetars,
but even for the ultra strong fields considered here
the decay time will be hundreds if not thousands of years.} Hence stationary
models of neutron stars in magnetic fields require a separation
of dynamical and dissipative time scales, encoded in an assumption
of infinite conductivity [magnetic fields ``frozen in'' and carried
with the fluid, a common
assumption in astrophysics \citep{alfv50}]. This assumption 
is exceedingly well justified for neutron star matter,
the Ohmic dissipation time scale being larger than the age of
the universe \citep{gold92}.
Cowling's theorem is thus effectively nullified.  
In addition, as shown by \cite{bona93},
the assumption of infinite conductivity 
leads to the requirement of uniform rotation, as well as
the relation 
\begin{equation}
A_t = -\Omega A_\phi +{\rm const} \label{arel}
\end{equation}
inside the
star, where the constant is
determined by the total electric charge of the star. 

Closure of the system of equations requires relations involving
some quantities appearing in the source equations (\ref{snu}-\ref{szeta})
and (\ref{sat}-\ref{saph}): the rest-frame energy density $e$ and pressure $p$,
and the components $J^t$ and $J^\phi$ of the electromagnetic 4-current.
For a uniformly rotating stationary star in a magnetic field,
local energy-momentum conservation ($T^{\alpha\beta}_{\ \ \ ;\beta}=0$) 
yields the equations of stationary equilibrium: \citep{bona93}
\begin{equation}
{1\over e+p}\; p_{,i} + \nu_{,i} - (\ln \Gamma)_{,i}
	- {1\over e+p}F_{i\alpha}J^\alpha=0, \label{hydroapp}
\end{equation}
in which $X_{,i}\equiv \partial X /\partial x^i$.
Derivatives with respect to the coordinates 
$x^i=(r,\theta)$ give the only nontrivial equations, due to the
symmetries of the problem. We recall that
$\nu\equiv(1/2)\ln (-g_{tt})$ here plays a role like that of the
gravitational potential in the Newtonian case, and that $\Gamma$ is
the Lorentz factor associated with the fluid's rotational velocity.
By analogy with the Newtonian case, the terms in equation (\ref{hydroapp})
may be thought of as (from left to right) the pressure force, gravitational
force, centrifugal force, and Lorentz force.

For a one-parameter equation of state, $e=e(n)$, $p=p(n)$, there
is a first integral of the first term in equation (\ref{hydroapp}):
\begin{equation}
h(n) = \int_0^n {1\over e(n') + p(n')}\,{dp\over dn'}\,dn',\label{heatfunc}
\end{equation}
where we will assume that $h(0)=0$ corresponds to the surface of the
star.\footnote{For example, at zero temperature 
and in chemical equilibrium, the
pressure and energy density are functions only
of the baryon density $n$. By virtue of the 
first law of thermodynamics, $h(n)$ is seen to be the logarithm of
the enthalpy per baryon: $h = \ln[(e+p)/ n] + \rm{constant}$.}
This, together with the adoption of a ``current function''
$f(A_\phi)$,
\begin{equation}
{1\over e+p}(J^\phi - \Omega J^t)=f(A_{\phi}), \label{current}
\end{equation} 
give the equations of hydrostatic equilibrium a first integral
\citep{bona93}:
\begin{equation}
h(r,\theta)+\nu(r,\theta) - \ln \Gamma(r,\theta)+M(r,\theta)
	 =C=\rm{constant},
\label{bern2}
\end{equation}
where the ``magnetic potential'' $M(r,\theta)$ is given by
\begin{equation}
M(r,\theta)=M(A_\phi(r,\theta))=- \int_0^{A_\phi (r,\theta)} f(x) dx.
\label{magpot}
\end{equation}
While one is free to choose the current function $f(A_\phi)$,
equation (\ref{current}) represents a significant restriction
on the form of the electromagnetic current that allows
the existence of stationary solutions.
The constant $C$ is determined by an input parameter, e.g. 
the density specified at some point in the star.

In summary, the formalism of stationary neutron stars in poloidal
magnetic fields with a one-parameter equation of state 
consists of a closed system of eleven variables [four metric
variables, energy density, pressure, two components 
of the electromagnetic potential, two components of the
electromagnetic current, and the ``heat function'' $h$ of equation
(\ref{heatfunc})]; eleven
equations [four Poisson equations (\ref{d1}--\ref{d4}) 
for the metric variables,
two Poisson equations (\ref{ateq}-\ref{apheq}) 
for the components of the electromagnetic
potential, the relation (\ref{arel})
between the components of the electromagnetic
potential, the equation of state, the relation (\ref{heatfunc}) between the
heat function and $e$ and $p$, the first integral (\ref{bern2}) of the 
equations of hydrostatic equilibrium, and
the restriction (\ref{current}) on the electromagnetic current]; three input
parameters (angular velocity, total electric charge, and
maximum density); and one input function [the $f(A_\phi)$ in
equation (\ref{current})].
 
\section{WHEN IS THE MAGNETIC FIELD TOO LARGE?}
\label{sec:himass}

As noted in the introduction, in this study we restrict
ourselves to the static case.
This involves a number of
simplifications, including the vanishing of $N^\phi$, $A_t$
and $J^t$, and the absence of surface charges
generally present on perfectly conducting rotating bodies.

\cite{bocq95}
present numerical calculations 
aimed at determining
``the maximum mass configuration among all static
magnetized models'' for several equations of state
and for the choice of constant current function $f(A_\phi)=f_0$.  
These static configurations are determined by two parameters, which
\cite{bocq95} took to be the central density and $f_0$.  They
considered sequences of constant magnetic dipole moment ${\cal M}$,
which is defined in terms of the asymptotic behavior of (the
orthonormal components of) the magnetic field,
\begin{equation}  
(2{\cal M}\cos\theta/r^3)=B_{(r)}|_{r\rightarrow\infty}=
(e^{2\nu-\zeta}/Gr^2\sin\theta) (\partial
A_\phi/\partial\theta)|_{r\rightarrow\infty}\,,
\label{magmom}
\end{equation}
and then determined the maximum mass for each value of ${\cal M}$. For
example, using the ``Pol2'' $\gamma=2$ polytropic equation of state
\citep{salg94}, they reported the maximum mass among all static
configurations to be $4.062{\rm M}_\odot$, with a magnetic moment of
$1.122 \times 10^{33}$ A m$^2$.  We have calculated, also using the
Pol2 EOS, a similar configuration, which is pictured in Figure
\ref{contour}, and may be compared with Figures 5 and 6 of
\cite{bocq95}.  We point out that in Figures \ref{contour} through
\ref{dnudr} that we use the same polytropic constant in the Pol2 EOS
as that employed by \cite{bocq95}.  However, in later figures and
in all tables, we have altered the polytropic constant to a more
realistic value so that the maximum mass of static configurations is
$2.0 {\rm M}_\odot$, and then scaled the results from the other groups
accordingly in these tables and figures.

\cite{bocq95} also note that ``For magnetic fields higher than [this
configuration], no stationary configuration can exist and the
numerical procedure\ldots fails to converge.'' Note that this is not a
question of {\em stability}---nothing is being claimed about the
stability of the high field configurations they achieve---the question
pertains to even the {\em existence} of stationary solutions.  The
fact that the maximum mass and magnetic moment are determined by the
failure of their code to converge leaves one wondering whether a
genuine physical limit has been reached, or whether the result simply
represents the failure of the numerical method to find solutions that
do in fact exist.

This situation is in contrast to the case of rotation: It is well
known that there is an upper limit to the angular velocity of
uniformly rotating stars.  Numerical codes that attempt to solve for
stellar configurations with uniform angular velocities above this
limit will fail to converge. However, there is also a well-defined
physical basis for this ``Keplerian'' limit, namely, mass
shedding.\footnote{In reality, the rotation may be more severely
limited by a gravitational instability to non-axisymmetric
perturbations. The instability might, however, be damped out by bulk and
shear viscous effects
(\cite{Lind77,Imam85,Frie86,Ipse89,Sawy89})
To the extent that the Keplerian limit represents a reasonable 
estimate of the upper limit of rotation, our analysis here
illuminates the stark contrasts that exist between the instabilites caused by 
rotation and magnetic fields.}

At the mass shedding limit, a fluid element at the
equatorial surface undergoes geodesic motion: it can remain in that
orbit without any pressure support from the star. 
Thus, it is not necessary to rely only on the failure of the code 
to converge to determine the mass shedding limit; one can
quantitatively test for geodesic motion 
at the equatorial surface.

The approach to the mass shedding limit can also be visualized
using the first integral of the equations of hydrostatic 
equilibrium---the relativistic generalization of the Bernoulli equation.
From equation (\ref{bern2}), the ``Bernoulli equation'' for a 
uniformly rotating star without a magnetic field is
\begin{equation}
h(r,\theta) + \nu(r,\theta) - \ln \Gamma(r,\theta)=C=\rm{constant}.
\label{bernrot}
\end{equation}
Figure \ref{potrotmult} demonstrates how the
Bernoulli equation can be used to visualize the approach to the mass
shedding limit. In this figure the dashed lines represent $\nu$,
the dot-dashed lines represent $-\ln \Gamma$, and the thick solid
lines represent $\nu -\ln\Gamma$, 
all as functions of a compactified radial coordinate
in the
equatorial plane. The dotted lines represent $C$, so that $h$ is given
by the distance between the dotted and thick solid lines. The
surface of the star ($h=0$) then corresponds to the intersection
of the dotted and thick solid lines. In the lower panel of Figure 
\ref{potrotmult},
which represents the mass shedding limit, the equatorial surface
of the star coincides with the maximum of $\nu -\ln\Gamma$.
For larger angular velocities, $C$ would be larger than this
maximum, and there would be no surface of the star at finite radius.

Similar plots can be made for stationary stars with magnetic fields.
For a nonrotating star in a poloidal magnetic field,
equation (\ref{bern2}) reduces to
\begin{equation}
h(r,\theta)+\nu(r,\theta) + M(r,\theta)
	 =C=\rm{constant}
\label{bernmag}.
\end{equation}
Now, in the equatorial plane, 
along which direction are magnetic forces exerted?
The answer depends on the direction of 
the magnetic field, which in turn
depends on the current distribution in the star.
From equation (\ref{current})
(with $\Omega$ and $J^t$ set to zero in the present static case),
the electric current density is proportional to 
$(e+p)$, so that in the 
typical case one would expect the
current to vanish on the surface of the star.
The current measured by a local observer
must also vanish on the axis of symmetry.
Hence the current $J^{(\phi)}$ measured by a local observer 
in the equatorial plane is
expected to have a structure that peaks somewhere inside the
star and vanishes at the origin and the surface, as shown in the left
panel of Figure \ref{radial} for the Pol2 EOS. 
Since the
magnetic field lines tend to circle around a point in the 
vicinity of the maximum current (see the right panel in 
Figure \ref{contour}), in the equatorial plane
the field reverses direction inside the star. 
Accordingly, the Lorentz force reverses direction inside
the star, as displayed in the right panel of Figure \ref{radial}.

The Lorentz force and the other ``forces'' acting in the equatorial
plane are derivatives of the quantities appearing in equation
(\ref{bernmag}). These quantities are plotted in Figure
\ref{potmagmult} for the Pol2 EOS, which is similar to Figure
\ref{potrotmult}, except that the dot-dashed lines now represent the
magnetic potential $M(r,\theta)$ in equation (\ref{bern2}), given in
this case by equation (\ref{magpot}) with $f={\rm constant}=f_0$.  In
the inner portion of the star, the Lorentz force behaves like a
centrifugal force, pushing outward and allowing the star to support
more mass. As seen in the lower panel of Figure \ref{potmagmult} (and
in the left panel of Figure \ref{contour}), this outward force can be
strong enough to cause the maximum of $h$ (and hence $e$) to be
off-center.  However, the reversal of the force at larger radii makes
the total ``potential'' confining.  Thus, in contrast to the case of
rotation, there is no mass shedding limit at the equatorial surface
that clearly indicates the nonexistence of stationary solutions and
explains the failure of the code to converge for large magnetic
fields.

If mass shedding does not occur, is there some other 
identifiable physical
cause that prevents the existence of stationary solutions
for sufficiently large magnetic fields? \cite{bocq95}
note that for sufficiently large fields, the total (fluid + magnetic)
stress tensor has a component on the symmetry axis that goes from 
being positive (pressure) to negative (tension), causing the star
to have a characteristic ``pinched'' shape 
(see the left panel of Figure \ref{contour}). 
This occurs when the ``magnetic pressure''
exceeds the fluid pressure. They note that in the largest mass configurations
they obtain, the ratio of the magnetic to fluid pressures at the
center of the star is of order unity. However, there are two reasons
that argue against considering this as 
a definitive physical reason preventing
the existence of stationary solutions. The first is a 
matter of principle: the equilibrium
of fluid elements relies on a balance between gravity and 
and the {\em gradients} of stresses; the absolute sign of the stresses
themselves is not of fundamental importance. 
The second reason is the
results observed in practice:
the value of the ratio of magnetic pressure to fluid pressure at
the center of the star reported by \cite{bocq95}, 
for the putative maximum mass configurations, varies quite noticeably
among different equations of state, and its value is not
predictable. This is not what one would
expect of a precise physical criterion for the nonexistence of
stationary solutions.

For Newtonian stars, \cite{chan53} identified gravitational
binding, or a negative
total energy (excluding rest mass), as a 
necessary criterion for the dynamical
stability of equilibrium solutions. For nonrotating polytropes, they
used a generalization of the virial theorem that accounts for magnetic
fields to show that a negative total energy requires that the
magnitude of the gravitational potential energy must be greater than the
magnetic field energy. 

One might wonder whether a criterion of gravitational binding
could represent a physical upper limit on stellar magnetic fields 
in the relativistic case as well. While total gravitational and
magnetic energies are not well defined in relativity, in some
sense the gravitational mass $M$ contains all forms of ``energy,''
including ``magnetic energy.'' 
Since the total baryon number [or total baryon mass $M_B=$ (baryon mass)
$\times$ (baryon number)] is a well defined quantity, one might
consider using the relation $M-M_B<0$ as a criterion for gravitational
binding. As it turns out, the putative maximum mass and maximum
field configurations reported by \cite{bocq95} are still
``gravitationally bound'' by this criterion. 
The question remains:
As the influence of the
magnetic field
increases, is this apparent transition from existence to nonexistence 
of static solutions, occuring at finite ``gravitational binding,'' a
physical result or a numerical artifact?

In order to explore these questions we have searched for the maximum
mass in a different way than Bocquet et al. (1995).
As mentioned previously, \cite{bocq95}
computed sequences of constant magnetic moment ${\cal M}$ by suitably
adjusting the central log-enthalpy $h_c$ and the value $f_0$ of the
(constant) current function $f$. For each value of ${\cal M}$ they
determined the maximum mass. The overall maximum mass was obtained
with the largest value of ${\cal M}$ for which convergence could be
achieved. In contrast, we have chosen a more direct means of exploring
parameter space. For each value of maximum log-enthalpy $h_{\rm max}$
(which for large fields will {\em not} coincide with $h_c$), we found
the largest values of $f_0$ for which the code converged.  Specifying
$h_{\rm max}$ instead of $h_c$ allows for the possibility of vanishing
density at the origin, i.e.  toroidal configurations.

A close inspection of the forces in configurations 
near the failure to converge
reveals
an apparent
physical cause,
for the choice of constant current function,
of the failure to find stationary solutions
for sufficiently large magnetic fields: 
{\em When a sufficient quantity of matter has been pushed off-center by 
magnetic forces that gravitational forces  
begin to point radially outward in the equatorial plane, 
a topological change to a toroidal
configuration ensues.} 
This is illustrated by
Figure \ref{runaway} for the Pol2 EOS, which 
shows the late stages of iteration of a configuration
with a value of $f_0$ that is slightly too large for convergence.
We emphasize that the stages depicted in this figure are not
valid stationary solutions to the Einstein equations; neither do they
represent a true evolution. Nevertheless, the sequence is suggestive
of possible dynamical outcomes: a transition to a toroidal topology,
expansion of the torus to large radii, and increasing compactification
of the toroidal configuration of matter. 
As the iterations proceed, the outward pointing gravitational force
(positive $-\partial\nu/\partial s$ or $-\partial \nu / \partial r$)
becomes more and more pronounced. The central evacuation begins,
however, with a tiny outward gravitational force at some radius.

Since we were unable to find any convergent toroidal solutions for the
case of a constant current function $f$, we can therefore identify a
quantitative criterion for the boundary of existence of stationary
solutions in this case: this
boundary is characterized by $\partial\nu/\partial r = 0$ at some
off-center location in the equatorial plane. By symmetry, it is always
the case that $\partial\nu/\partial r = 0$ at the origin; furthermore,
the positive-semidefiniteness of energy density and pressure ensure
that $\partial^2\nu/\partial r^2 \ge 0$ at the origin (this can be
verified from the Green's function expansions of the metric functions
presented in Appendix A).  This means that the critical condition
$\partial\nu/\partial r < 0$ cannot occur at the origin or an
infinitesimal region surrounding it, but our numerical calculations
indicate that it first occurs very close to the origin.

These matters are illustrated in Figure \ref{dnudr} with the Pol2 EOS,
which shows the radial profile of the gravitational force for
increasing values of $f_0$---at a particular value of maximum
density---including the largest value of $f_0$ for which convergence
was achieved. In the upper panel, a flattening of the gravitational
force near the center with increasing magnetic field is apparent.  The
insets in the upper panel show the region near the origin.  The lower
left inset shows $-\partial\nu /\partial r$ as computed with a formula
obtained by differentiating equation (\ref{nugreen}).  As required by
the condition $\partial^2\nu/\partial r^2 \ge 0$ at the origin,
according to this formula the first off-origin grid point has negative
value of $-\partial\nu /\partial r$. For the largest value of $f_0$,
the next grid point is less negative, but still does not reach our
asserted condition $-\partial\nu /\partial r=0$. However, the upper
right inset in the upper panel shows values of $-\partial\nu /\partial
r$ near the origin computed from a the centered finite difference
formula, which is closer to what the discretized configuration
actually ``feels.'' Computed in this way, the first off-origin grid
point has a barely positive value of $-\partial\nu /\partial r$ for
the largest value of $f_0$ for which convergence was achieved.  The
lower panel of Figure \ref{dnudr} shows the gravitational force for
the largest value of $f_0$, for three different grid resolutions. The
insets of this panel again compare the ``analytic'' and ``numerical''
derivatives. These computations show that with increasing resolution,
the ``analytic'' derivative near the origin gets closer and closer to
the critical condition $-\partial\nu /\partial r=0$. In practice,
however, the maximum mass and magnetic field configuration for a given
stellar maximum density can be identified by the appearance of a small
positive value of $-\partial\nu /\partial r$, as computed with a
centered finite difference.

For the values of maximum
density we studied, and for a constant current function $f$, we
did not find any stationary toroidal solutions. The toroidal configurations
continued to expand in radius and 
compress into thinner and thinner rings until the region  covered by matter
consisted of only a few gridpoints, at which point the code would fail.
A determination of the outcome of the evolution of such
configurations---whether to a stationary solution characterized by a different
current function, dispersal to infinity, or even the formation of a
toroidal event horizon \citep{shap95}---would
appear to require a fully relativistic evolution code.

It remains to be seen if these results---namely, the prescription for
determining when magnetic forces are too strong for the existence
of static configurations, and the lack of converged
toroidal solutions---will hold for more general current functions.

\section{SEQUENCES OF CONSTANT BARYON MASS AND CONSTANT MAGNETIC MOMENT }
\label{sec:constmb}
Our computations of static neutron stars with ultra-strong magnetic
fields determined by a constant current function are summarized in
Table \ref{tbl-25} and Figures \ref{mr1} and \ref{mr2}.
Figure \ref{mr1} contains some of the older EOSs employed by
\cite{bocq95}.  These include BJI, which is model IH of \cite{beth74}
[we derived our table from that listed in \cite{malo75}], and PandN,
which is from \cite{pand71}.  Results for more recent EOSs are
displayed in Figure \ref{mr2}.  These include Akmal, which is from
\cite{akmal98} and is based on a potential model description of dense
matter and represents the most complete study to date in which
many-body and special relativistic corrections are incorporated.  PCL
is taken from \cite{pcl95}, and is based on a relativistic
field-theoretical description of dense matter starting from the
Lagrangian proposed by \cite{zm90}.  This  approach
easily allows for the inclusion of additional softening components:
the case in which hyperons are present is labelled PCLhyp.  
For all EOSs except Pol2, we employed the EOS of \cite{baym71a} and
\cite{baym71b} at densities below about 1/2 the nuclear saturation density.

The rationale for exploring a wide variety of EOSs, even some that are
relatively outdated, is two-fold. First, it provides contrasts among
widely different theoretical paradigms. Second, it illuminates general
relationships that exist between the pressure-density relation and the
macroscopic properties of the star such as the maximum mass and the
radius.

In these figures the lower thick solid curves show the gravitational
mass $M$ as a function of radius for static stars without magnetic
fields, which are spherical.  The upper thick solid curves show the
outer boundaries beyond which no static solutions were found (see
\S\ref{sec:himass}). The lighter solid curves are sequences of
constant baryon mass $M_B$ (and varying magnetic moment $\cal M$),
while the dotted curves are sequences of constant magnetic moment (and
varying baryon mass).  The lighter shaded regions indicate
configurations in which the magnetic field is sufficiently strong that
the maximum density is off-center. The small slivers of darkly shaded
regions towards the left sides of the plots indicate solutions which
are gravitationally unbound ($M-M_B>0$); these are expected to be
dynamically unstable.

As with rotation, magnetic fields allow neutron stars with a
particular EOS and baryon number to have larger masses and equatorial
radii compared to the field-free case. In Figures \ref{mr1} and
\ref{mr2}, the configurations of maximum mass that can be reached by
uniform rotation (without magnetic fields) are marked with an
``X''. For all the EOSs displayed, our result for the maximum mass
attainable with a magnetic field governed by a constant current
function is noticeably larger than that attained by rotation. This
contrasts with the results reported by \cite{bocq95} for the maximum
mass attainable with magnetic fields, shown in Figure \ref{mr1} with
crosses.  Out of the three EOSs we have in common, only in the case of
the polytropic EOS do \cite{bocq95} obtain a significantly larger mass
with magnetic fields than with rotation.

In the absence of significant accretion, constant baryon mass
sequences are of interest as potential evolutionary paths. This is
well motivated in the case of uniformly rotating non-magnetic stars:
As angular momentum is slowly dissipated by gravitational radiation,
the star moves along a sequence of constant baryon mass until it
either stops rotating (for ``normal'' sequences terminating on the
spherical mass vs. radius curve) or collapses to a black hole [for
``supramassive'' sequences which exist solely by virtue of rotation;
stars on such sequences may exhibit the interesting phenomenon of
spin-up during angular momentum loss just before collapse to black
hole \citep{cook92,cook94a,cook94b,salg94}]. These scenarios are well
motivated because of the expectation that once viscosity brings a
dynamically stable star into uniform rotation, it will not
spontaneously begin to differentially rotate as it loses angular
momentum.

On the other hand, representing evolutionary sequences by the constant
baryon mass sequences for magnetized stars pictured in Figures
\ref{mr1} and \ref{mr2} is an oversimplification.   Even though a star's
magnetic
field
will slowly 
(on dynamical time scales) decay due to Hall drift and ambipolar diffusion,
there is no guarantee that 
the star's configuration
will proceed along the paths pictured
in the figures. This is because the case of magnetic fields is more analogous
to differential rotation rather than uniform rotation, the necessary
choice of a current function in the magnetic case [see equation (\ref{current}) and surrounding discussion] corresponding to the choice of a rotation
law in the case of differential rotation. As a star's magnetic field
decays, it is not obvious that its current function will remain the 
same.\footnote{Note that constant current function is {\em not} analogous
to uniform rotation. Formally, the analogue to uniform rotation would be
uniform magnetic vector potential, which of course means no magnetic field.}
Perhaps the study of several different current functions could shed light
on probable evolutionary sequences. For example, an analysis of how the
mass varies with the functional form of the current function (at fixed
baryon mass and magnetic moment) could give an idea of how the slow evolution
with magnetic field decay might proceed. There is of course no principle of
``conservation of magnetic moment,'' but since the time scale for magnetic
field decay is slow, 
this procedure seems
like a plausible
opening exploration.

In connection with the constant baryon mass sequences, we here comment
on a curious feature in Figures \ref{mr1} and \ref{mr2}.  For the
potential model EOSs BJI, PandN and Akmal, the topology of these
sequences near the maximum spherically symmetric star appears to be
different than for the polytropic or relativistic field theoretical
models PCL and PCLhyp.  In the former, there is a minimum in the
baryon mass above the spherical, non-magnetized, sequence.  To
determine if this feature was related to the possible acausal
behavior of potential models at high density, we modified the
PandN EOS to go over to a causality limit EOS when necessary,
but found that the topology was unaltered. 
Instead, the effect appears to be related to the fact
that all forms of energy contribute to the magnetic field: While the
Newtonian intuition (and the relativstic behavior at lower densities)
is that magnetic fields always increase the gravitational mass of a
star of given baryon number, the fact that the energy density in
magnetic fields can be a nontrivial source of gravitation means that
this self-gravitating tendency can compete with the tendency of the
Lorenz force to help support the star. Perhaps this occurs near the
spherical maximum mass in the case of the potential model EOSs, with
the result that configurations near the spherical maximum mass have
nowhere to go but down in gravitational mass when magnetic fields
(governed by a constant current function) are added. If a star did
preserve its current function and follow a supramassive constant
baryon mass sequence possessing such a gap, upon reaching the minimum
mass of the sequence it could catastrophically collapse to a spherical
neutron star rather than a black hole. While the possibility is
admittedly remote, this could be a novel form of energy release in a
baryon-free environment, giving rise to a mini-$\gamma$-ray burst.

Except for the spherical stars, those with no magnetization indicated
by the lower solid line in these figures, the stability of the
configurations has not been studied.  For spherical stars, it is
well-known that the configurations with larger radii than the maximum
mass configuration are stable.  One might speculate that constant
baryon mass sequences which terminate on the stable side of the
spherical $M$ vs. $R$ curve are stable, while those ending on the
unstable side would be unstable, and that for supramassive constant
baryon mass sequences the minimum mass marks a change from stability
to instability.  But this remains to be determined.  As with
differentially rotating stars, it is necessary to do a normal mode
analysis, or even a fully relativistic evolution; see
\cite{baum00}. Incidentally, it is interesting to note that these
authors find a differentially rotating configuration with a ``red
blood cell'' shape similar to our extreme magnetic configurations, and
that this configuration is dynamically stable. Of course, their
configuration was not subject to MHD instabilities that may come into
play \citep{spru99a,spru99b}.

\section{SUMMARY AND OUTLOOK}
\label{sec:sumout}

In summary, we present a method of computing the
structure of axisymmteric relativistic stars that combines
elements of previous approaches, and report tests of our
code. A quantitative method of determining
the outer envelope (in the mass vs. radius plane) 
of configurations attainable with poloidal
magnetic fields governed by a constant ``current function'' [see equation
(\ref{current})] has been found:
magnetic fields are too large for static configurations 
to exist when the magnetic force 
pushes a sufficient amount of mass off-center that the gravitational
force points outward near the origin in the equatorial plane. 
(In our coordinates an outward gravitational force corresponds to 
$-\partial \nu/\partial r>0$.) 
We obtain larger masses of neutron stars in
ultra-strong magnetic fields than have been reported previously
for various equations of state, and performed computations with
three representative modern EOSs. Sequences of constant baryon
mass and constant magnetic moment are displayed. For all EOSs studied,
the maximum attainable mass of static stars
with a magnetic field determined by a constant current function 
is noticeably larger than that attainable with
uniform rotation and no magnetic field. 

The results presented here are only an initial step in exploring
possible configurations of neutron stars with strong magnetic fields.
As we mention below, configurations with azimuthal field components
will be of physical interest, which implies that three-dimensional
geometries should also be considered. But even with attention
restricted to poloidal fields, we have only scratched the surface of
possible configurations, as we have only considered a single current
function. \cite{bocq95} make brief mention of computations with a few
other current functions.  We have performed a handful of exploratory
computations using a polytropic EOS and other current functions and
have found some toroidal solutions.  These toroidal
solutions were not attainable with the computational approach
of \cite{bocq95}, since their method involved the specification
of a finite density at the center of the star.  
In the case of toroidal configurations,
the simple condition determining the boundary of existence of 
static configurations will
have to be generalized, since there is no matter at the center.  These
explorations will be reported in detail elsewhere.

Our work here has focused on the effects magnetic fields have on
general relativistic structure, ignoring the effects of intense
magnetic fields on the EOSs.  Recently, the direct effects of magnetic
fields on the EOS have also been investigated
\citep{cha96,cha97,yz99,bpl00}.  Substantial effects on the EOS above
nuclear saturation densities are generally produced by fields in
excess of $10^{18}$ G, which is of the order of the maximum central
field strengths found in this paper.  The generic effects on the EOS
include softening due to Landau quantization, which is, however,
overwhelmed by stiffening due to the incorporation of the magnetic
moments of the various particles in neutron star matter \citep{bpl00}.
(Note that the important $B^2/8\pi$ term is already included in our study.)
Work is in progress (Cardall et al. 2000, in preparation) to provide
fully self-consistent calculations of neutron star structure,
in which the direct effects of magnetic fields on the EOS will be included
in addition to the structural effects considered in this work.

An issue which we defer to future work is the question of stability.
For systems governed by a finite number of parameters, a generalization
\citep{sork82}
of the familiar one-dimensional turning point method can be employed;
see \cite{frie88} for an application to uniformly rotating relativistic
stars. However, as with differentially rotating stars,
this generalized turning point method is not really applicable in the
present case.
This is because the need to specify a current function (or rotation law
in the case of differential rotation) means that defining a particular
configuration requires the specification of an {\em infinite} number of 
parameters.

Another issue that needs further explication before the physical
relevance of the results presented here can be fully assessed 
is the generation of magnetic fields. We have mentioned the
mechanism of \cite{dunc92}, the generation of magnetic fields
during the smoothing of differential rotation. However, 
the azimuthal dragging of field lines by differential rotation
leads to nonvanishing azimuthal field components, 
in constrast to the poloidal fields
studied here. It is not clear whether fields with azimuthal
components would evolve into poloidal configurations, or whether
there are mechanisms to directly generate strong poloidal fields.
It would be of interest to explore the possibility of finding
stationary solutions with toroidal magnetic fields, and in three
dimensions. While this would involve more nonzero metric components,
perhaps methods similar to those employed in this paper could be
employed; see \cite{bona98}.

We wish to thank E. Gourgoulhon for helpful communications concerning 
the calculations of \cite{bocq95}.
We are grateful to Dany Page and Ralph Wijers for their help in the
preparation of Table 1. Research support from DOE grants
FG02-87ER40317 (for CYC and JML) and FG02-88ER-40388 (for MP) are
gratefully acknowledged.

\appendix

\section{NUMERICAL PROCEDURES AND TESTS OF THE CODE}
\label{sec:tests}

\cite{bona93} have developed a ``spectral method'' to 
solve equations (\ref{d1}-\ref{d4}). This method involves 
expanding the solution on a set of basis functions having
the analytical properties of the solution. They also 
try to choose basis functions for which there exist fast
transform algorithms. Their method involves the use of two
grids, an ``inner'' region including the origin and an
``outer'' region that reaches to infinity. Instead of this
``spectral method,'' we have chosen to solve the Poisson equations
using Green's functions, similar to the methods of
\cite{koma89} and \cite{cook92,cook94a,cook94b}. 
It is convenient to
compactify the radial domain $0\le r\le \infty$
to $0\le s \le 1$ via the
change of variables
\begin{equation}
r=R\left({s\over 1-s}\right), \label{radvar}
\end{equation}
where $R$ is some length scale.
Taking into account the
azimuthal and equatorial symmetries, and imposing
the boundary conditions [all metric functions finite at the
origin; $(\nu ,\, N^\phi, \, \zeta)|_{r\rightarrow \infty}\rightarrow
0$, $G|_{r\rightarrow\infty}\rightarrow 1$], we find equations 
(\ref{d1}-\ref{d4}) to yield

{\samepage
\begin{eqnarray}
\nu(s,\theta)&=&-\sum_{n=0}^{\infty} P_{2n}(\cos\theta)\times \nonumber \\
	& &\left[\left(
	{1-s \over s}\right)^{2n+1} \int_0^s {ds'\, s'^{2n}
	\over (1-s')^{2n+2} } \int_0^{\pi/2} d{\theta}'\,
	\sin\theta' P_{2n}(\cos\theta')\,
	\tilde\sigma_\nu (s',\theta') \right. \nonumber \\ 
 & & + \left. \left({s\over 1-s}\right)^{2n} \int_s^1 {ds'\,
	(1-s')^{2n-1} \over s'^{2n+1} } \int_0^{\pi/2} d\theta' \sin\theta'\,
	P_{2n}(\cos\theta')\, \tilde\sigma_\nu (s',\theta')
	\right], \label{nugreen}\\
\nonumber \\
N^\phi(s,\theta) &= &-{1\over R}\sum_{n=1}^\infty 
	{P_{2n-1}^1 (\cos\theta) \over
	2n (2n-1) \sin\theta}\times \nonumber \\
	& & \left[\left({1-s \over s}\right)^{2n+1}
	\int_0^s {ds'\, s'^{2n-1} \over (1-s')^{2n+1} } \int_0^{\pi/2}
	d\theta' \, \sin\theta' P^1_{2n-1}(\cos\theta')\,
	\tilde\sigma_{\tilde N^\phi}(s',\theta')\right.+ \nonumber \\
 & & \left.\left({s\over 1-s}\right)^{2n-2}\int_s^1 {ds'\, (1-s')^{2n-2}
        \over s'^{2n}}\int_0^{\pi/2} d\theta'\, \sin\theta' 
	P^1_{2n-1}(\cos\theta')\, \tilde\sigma_{\tilde N^\phi}(s',\theta')
	\right], \\
\nonumber \\
G(s,\theta)&=&1-{2\over \pi} \sum_{n=1}^\infty {\sin [(2n-1)\theta]
	\over (2n-1)\sin \theta}\times\nonumber \\
	& & \left[\left({1-s\over s}\right)^{2n}
	\int_0^s {ds'\, s'^{2n-1} \over (1-s')^{2n+1}} \int_0^{\pi/2}
	d\theta'
	\sin[(2n-1)\theta ']\, \tilde\sigma_{\tilde G}(s',\theta ')\right.
	\nonumber \\
	& & + \left. \left({s\over 1-s}\right)^{2n-2} \int_s^1
	{ds' \, (1-s')^{2n-3} \over s'^{2n-1}} \int_0^1 d\theta'\,
	\sin [(2n-1)\theta ']\,\tilde\sigma_{\tilde G} (s',\theta ')\right],\\
\nonumber \\
\zeta(s,\theta)&=&{2\over \pi}\left[ \ln r(s) \int_0^s{ds'\over (1-s')^2}
	\int_0^{\pi/2} d\theta' \tilde \sigma_\zeta(s',\theta')\right. 
	\nonumber \\
	& & \left. +
	\int_s^1 {ds'\over (1-s')^2} \ln r(s') \int_0^{\pi/2}
	d\theta' \tilde \sigma_\zeta(s',\theta')\right] \nonumber \\
	& &-{2\over \pi}\sum_{n=1}^\infty {\cos (2n\theta)
	\over 2n} \left[\left({1-s \over s}\right)^{2n} \int_0^s
	{ds'\, s'^{2n}\over (1-s')^{2n+2}}\int_0^{\pi/2} d\theta '\,
	{\cos (2n\theta ')}\, \tilde\sigma_\zeta
	(s',\theta ')\right. \nonumber \\ 
	& & + \left. \left({s\over 1-s}\right)^{2n} \int_s^1
	{ds'\, (1-s')^{2n-2}\over s'^{2n}} \int_0^1 d\theta '
	{\cos (2n\theta ')}\, \tilde\sigma_\zeta
	(s',\theta ')\right],\label{zeta}
\end{eqnarray} 
}
where
\begin{eqnarray}
 \tilde\sigma_\nu (s,\theta)&=&r^2 \sigma_\nu (s,\theta), \\
\tilde\sigma_{\tilde N^\phi}(s,\theta)&=&r^2 \sigma_{\tilde N^\phi}
	(s,\theta), \\
\tilde\sigma_{\tilde G} (s,\mu)&=&r \sigma_{\tilde G} (s,\theta),
\end{eqnarray}
in which the quantities on the right hand side are given by
equations (\ref{snu}-\ref{sg}). The symbols $P_n(x)$ and $P_n^m(x)$ denote the 
Legendre polynomial and the associated Legendre function, respectively. 
The source $\tilde\sigma_\zeta$ requires special consideration.
The Green's function of the 2D Laplacian has a $\ln r$ term, as
is apparent from equation (\ref{zeta}). 
This term must vanish as $r\rightarrow \infty$ in order to have
vanishing boundary conditions at infinity. This is not a problem
for equation (\ref{d3}) for $\tilde G$, since the $\sin\theta$ 
factor in the source $\sigma_{\tilde G}$
[see equation (\ref{sg})] guarantees that the
$\ln r$ term vanishes everywhere. However, there is no such factor in 
the source $\sigma_\zeta$ [see equation (\ref{d4})]. A genuine
solution to equation (\ref{d4}) 
satisifying the boundary condition
$\zeta|_{r\rightarrow\infty}\rightarrow 0$ will have a
$\ln r$ term that vanishes as $r\rightarrow \infty$; but in the
intermediate steps of an iterative procedure to solve the
nonlinear equations there is no guarantee that this will be
so, leading to a potential instability.

\cite{bona93} have a resolution of this difficulty which
we adopt here. In order for the $\ln r$ term to
vanish as $r\rightarrow \infty$ it is necessary that 
\begin{equation}
\int_0^\infty \int_0^{2\pi} \sigma_\zeta(r,\theta)\, r\,
	dr\, d\theta =0.
\end{equation}  
This condition is called the ``virial theorem'' by \cite{bona93}.
In terms of the variable $s$ it can be written as
\begin{equation}
\int_0^1 {ds\, s\over (1-s)^3}\, \sigma_{\zeta,0}(s)=0, \label{vir0}
\end{equation}
where 
\begin{equation}
\sigma_{\zeta,0}(s)=\int_0^{2\pi} d\theta\, {\sigma_\zeta (s,\theta)}.
\end{equation} 
The trick is to divide the source $\sigma_\zeta$ into two pieces.
One piece, $\sigma^m_\zeta$, contains the ``matter terms'' 
(those involving components of the stress-energy tensor); 
the other piece, $\sigma^f_\zeta$,
contains the ``field terms,'' those involving only the metric variables.
The virial theorem can then be written
\begin{equation}
\int_0^1 {ds\, s\over (1-s)^3}\, \sigma^m_{\zeta,0}(s)=
-\int_0^1 {ds\, s\over (1-s)^3}\, \sigma^f_{\zeta,0}(s).\label{vir}
\end{equation}
This equation will be satisfied for the actual solution to the
Einstein equations, but will not be satisfied in the intermediate
steps of the iteration procedure. 
To avoid the potential logarithmic divergence
associated with the failure to satisfy equation (\ref{vir}),
the source $\sigma_\zeta$ is replaced by $\sigma^m +\lambda\, \sigma^f$,
where 
\begin{equation}
\lambda= 
-{\left[\int_0^1 {ds\, s\over (1-s)^3} \sigma^m_{\zeta,0}(s) \right]
/
\left[\int_0^1 {ds\, s\over (1-s)^3} \sigma^f_{\zeta,0}(s)\right]}.
\label{lambda}
\end{equation}
In this way equation (\ref{vir0}) is satisfied at each step of the
iteration---avoiding the potential logarithmic singularity---but 
with $\lambda\ne 1$ in the intermediate steps. At the
end of the iteration process $\lambda$ must approach 1 for the
computed metric functions to represent
a valid solution to the Einstein equations. 
Finally, $\tilde\sigma_\zeta$ in 
equation (\ref{zeta}) is given by $\tilde\sigma_\zeta=R\, r\, (\sigma^m_\zeta
+\lambda \sigma^f_\zeta)$.

It is convenient to know ahead of time the location of the equatorial
surface on the grid. To achieve this we employ a scheme like that
\cite{bona93} use to divide their computational domain into ``inner''
and ``outer'' grids. Specifically, we specify that the equatorial surface
be located at the radial position $s=0.5$. From equation (\ref{radvar}),
this makes the equatorial radius equal to $R$. Since $R$ is some chosen
constant, this involves a nonstandard system of units. Operationally,
the value of Newton's constant $G_N$ is adjusted at each iteration in such a
way that $s=0.5$ does indeed correspond to the equatorial radius of
the neutron star, identified by $h=0$, where $h$ is defined by equation
(\ref{heatfunc}). The physical value of the equatorial radius scales
as $\sqrt{G_N}$; other physical quantities also involve various powers
of $G_N$.

The scheme is implemented as follows. As with $\sigma_\zeta$ described
above, the source term $\sigma_\nu$ is divided into two parts, with
the ``matter part'' $\sigma_\nu^m$ containing source terms deriving from the
stress-energy tensor, and the ``field part'' $\sigma_\nu^f$ containing
terms involving derivatives of the metric variables. The Poisson equation
for $\nu$ is solved in two parts: $\Delta_3\nu^m = \hat \sigma_\nu^m$,
where $\hat\sigma_\nu^m = \sigma_\nu^m/G_N$; and $\Delta_3\nu^f =
\sigma_\nu^f$. The full value of $\nu$ is then $\nu=G_N\nu^m +\nu^f$.
The demand that $h$ vanish at $s=s_*=0.5$ in the equatorial plane, together
with equation (\ref{bern2}), yields the appropriate value of Newton's
constant at each iteration:
\begin{equation}
G_N = {{\left(h + \nu^f - \ln\Gamma+ M\right)|_{s=
	s_{\rm max}}-
\left(\nu^f 
	- \ln\Gamma
	+M\right)|_{s=s_*}} 
	\over \nu^m|_{s=s_*}-
	\nu^m|_{s=s_{\rm max}} }.
\end{equation}
In this expression, 
$h|_{s=s_{\rm max}}$ 
is an input parameter
that, via equation (\ref{heatfunc}) and the equation of state, specifies
the maximum density in the equatorial plane. This is not necessarily the
central density; its location $s_{\rm max}$ must be determined at each
iteration. Specifying the maximum density while allowing its location
to ``float'' allows for the possibility of toroidal configurations, 
a possibility excluded by the method of \cite{bona93} and \cite{bocq95}. 

In order to test the ability of our code to solve the Einstein
equations for axisymmetric configurations, we have studied uniformly
rotating configurations with a poltropic EOS and two 
tabulated ``realistic'' EOS from the literature. Physical
characteristics of the maximum mass configurations for both nonrotating
and rotating stars are listed in Table \ref{tbl-1}, and compared
with the results of two other groups. While there is excellent
agreement across the board, the agreement is particularly good in the
polytropic case. Slightly larger differences in the case of the tabulated 
EOS are well-attested in the literature, resulting from different 
methods of interpolation, matching between different density regimes, etc.
We tried two methods of determining the ``heat function'' $h$ from the
tabulated EOS: 1) direct integration of equation (\ref{heatfunc}),
and 2) use of an analytic formula derived from equation (\ref{heatfunc})
with the first law of thermodynamics. \cite{bocq95} used
the analytic formula, and when employing this method we obtained
closer agreement with the results of
those authors. However, we achieved smaller values of $|1-\lambda|$
(indicating better solutions) when constructing $h$ by direct integration.
This was apparently the method employed by \cite{cook94b}, as our results
with this method agree more closely with theirs. Our calculations underlying
the entries in Table \ref{tbl-1}---and, indeed, our calculations 
with tabulated EOS
reported throughout this work---employed the construction of $h$ by
direct integration.

In our experience the calculations of rotating configurations
with tabulated EOS had a
tendency towards numerical instability for large angular velocities,
manifested  as growing oscillations in various quantities
(as opposed to the monotonic runaway occurring past the Keplerian
limit). We found that this instability could be controlled by updating
the metric variable $N^\phi$ only once every $n$ iterations, where
we took $n$ as high as 15 when approaching the Keplerian limit. This
procedure was not necessary with the polytropic EOS we used.

We also tested our code's ability to reliably compute 
static neutron star configurations with large magnetic fields.
As reported in the main text, our results for the maximum mass
of neutron stars with constant current function $f_0$ are larger
than those reported by \cite{bocq95}. Hence comparison of these
maximum mass configurations does not really constitute a verification
of our code. However, \cite{bocq95} also presented results for 
the maximum mass at a certain fixed low values of magnetic moment, and
comparison of these results provides a benchmark against which our
code can be checked. In addition, we can compute configurations
close to those
reported by \cite{bocq95} as the maximum mass, 
and make a comparison---even though these are not our maximum mass
configurations.

\begin{figure}
\epsscale{1.0}
\plotone{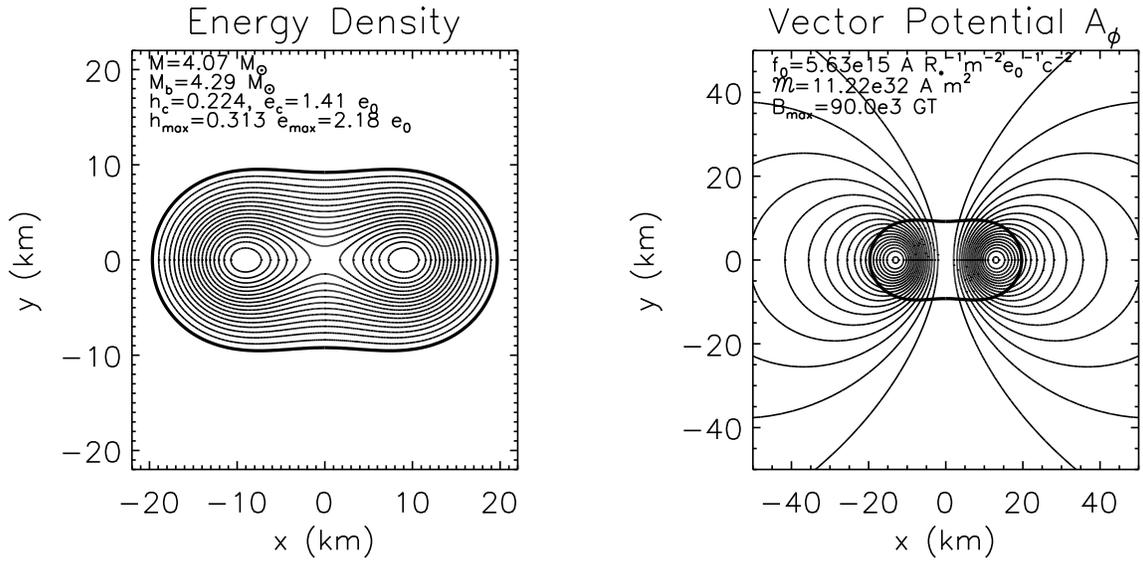}
\caption{Contour plots of the energy density and the 
electromagnetic
vector potential component $A_\phi$. Here $x=r\sin\theta$ and
$y=r\cos\theta$, where $r$ and $\theta$ are coordinates appearing in
equation (\ref{metric}); note that in these coordinates not even
distances in the equatorial plane constitute proper
distances. $e_0=1.66\times 10^{14}$ g cm$^{-3}$, and $R_*$ is the equatorial
radius. While contours of constant $A_\phi$ show the structure of the
magnetic field, their spacing as pictured here does not accurately
indicate magnetic field strength; the maximum magnetic field is
actually at the center of the star.  To allow a direct comparison,
with Figures 5 and 6 of \cite{bocq95}, we assumed their value of the
polytropic constant.  Table \ref{tbl-2} contains physical quantities
rescaled to reflect a more realistic value of this constant).}
\label{contour}
\end{figure}

\begin{figure}
\epsscale{0.6}
\plotone{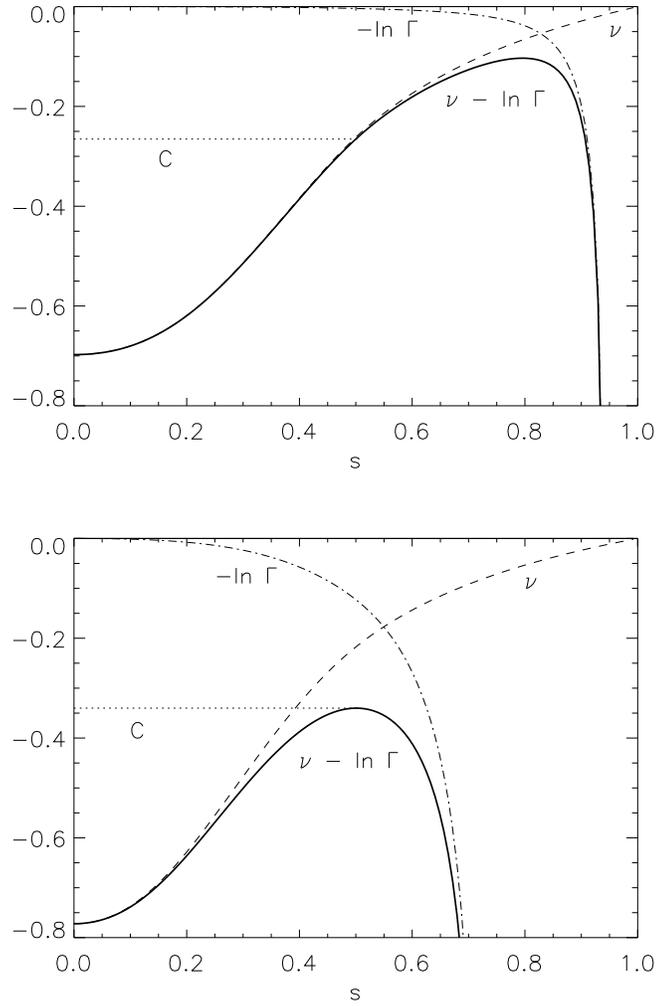}
\caption{Various 
quantities associated with the Bernoulli equation
(\ref{bernrot}) in the equatorial plane of a uniformly rotating star, as
a function of the compactified radial coordinate $s=r/(R_*+r)$, where $R_*$ is
the equatorial radius.
The upper panel shows a configuration with modest rotation, and 
the lower panel shows a configuration at the Keplerian limit.
See the text for discussion.} \label{potrotmult}
\end{figure}

\begin{figure}
\epsscale{1.0}
\plotone{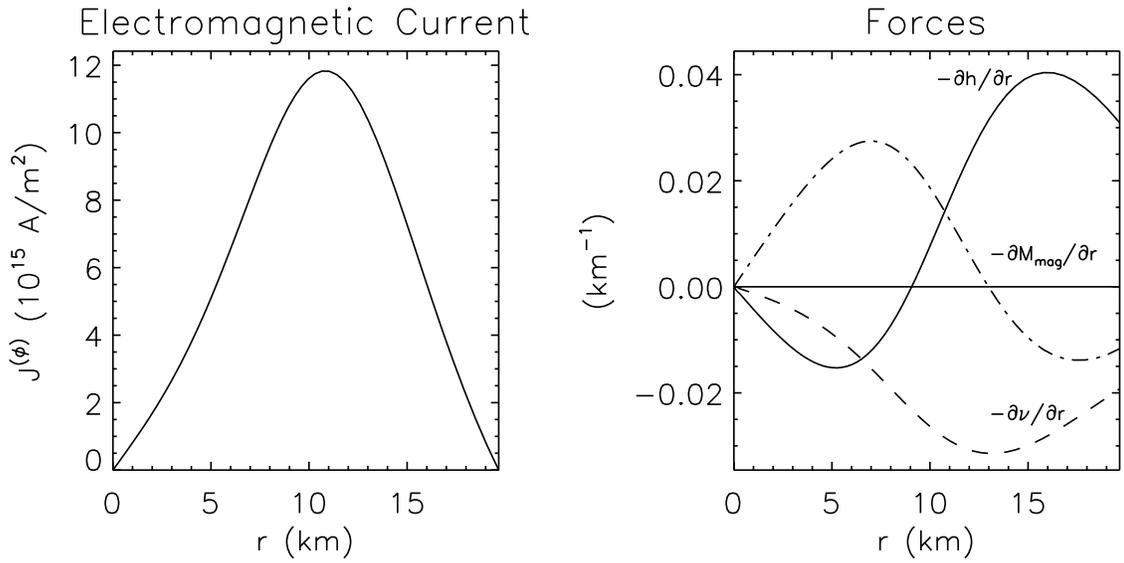}
\caption{The electromagnetic current and various 
``forces'' in the
equatorial plane for the configuration pictured in Figure
\ref{contour}.  The ``forces'' are derivatives of the terms in the
Bernoulli equation (\ref{bernmag}); here $M$ is the magnetic
potential.  The right boundaries correspond to the equatorial radius
$R_*$.  Note that the pressure force ($-\partial h/\partial r$) points
inward at lower radii; this is a consequence of the maximum density
being pushed off center (see the left panel of Figure \ref{contour})
due to the strong outward Lorentz force ($-\partial M/\partial
r$).  The Lorentz force also reverses sign, due to a reversal of
magnetic field direction in the equatorial plane (see the right panel
of Figure \ref{contour}). At $R_*$, the Lorentz force
works together with gravity to help confine the star, in contrast with
the centrifugal force in the case of rapid rotation.}\label{radial}
\end{figure}

\begin{figure}
\epsscale{0.6}
\plotone{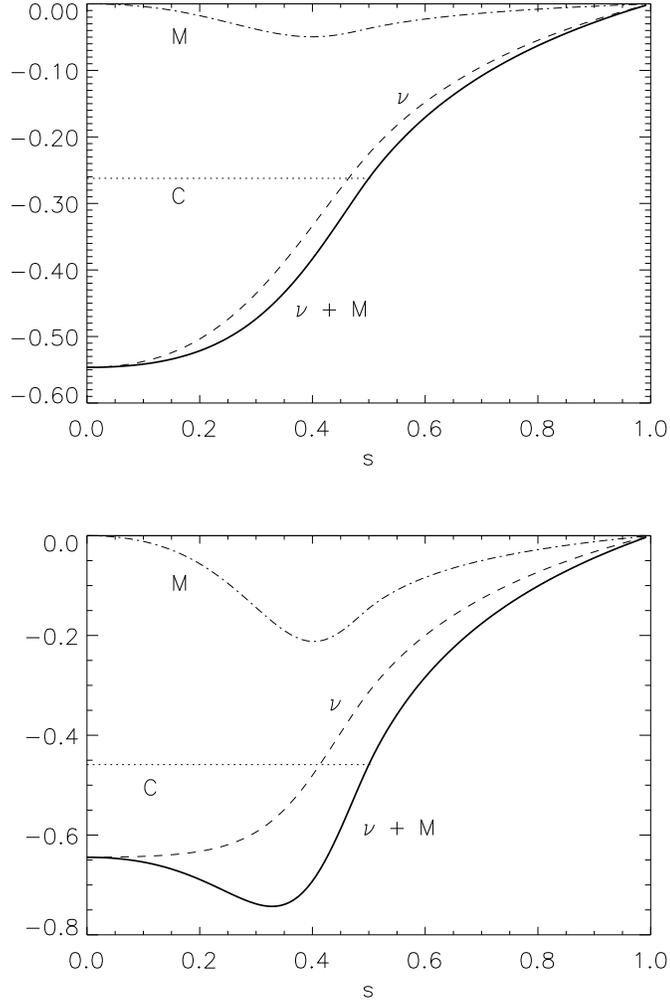}
\caption{Similar to Figure \ref{potrotmult}, 
but for the Bernoulli
equation (\ref{bernmag}) of a nonrotating star in a poloidal magnetic
field. $M$ is the magnetic potential.
The upper panel shows a configuration in which the magnetic field is
strong enough to modestly deform the star, while the lower panel
displays the configuration with the largest current function for which
convergence was achieved (for the same maximum density as the 
configuration in the upper panel).
When compared with Figure \ref{potrotmult}, the absence of mass shedding
is evident.}\label{potmagmult}
\end{figure}

\begin{figure}
\epsscale{1.0}
\plotone{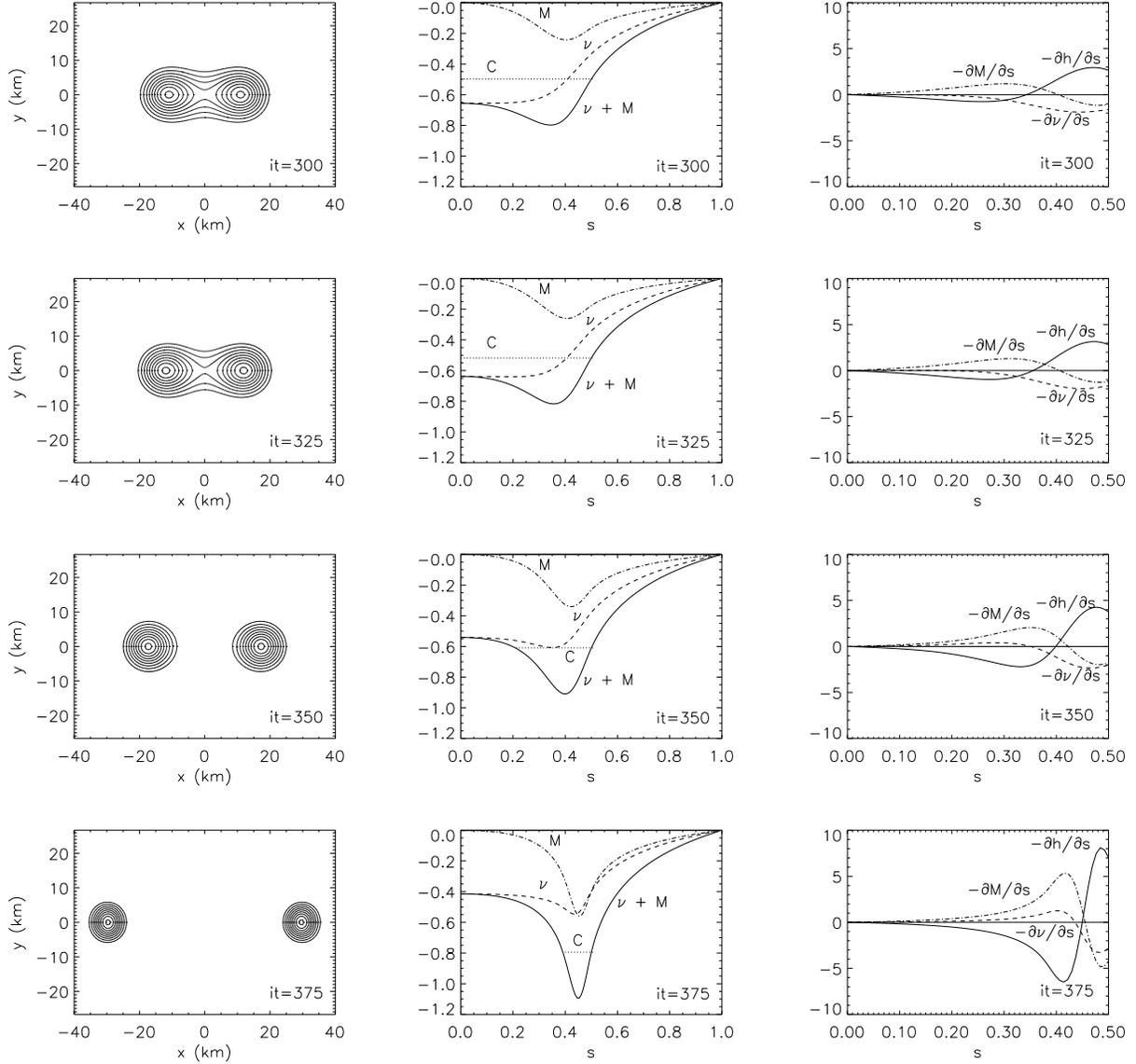}
\caption{Late stages of iteration of a 
non-converging configuration (as
a solution of the Einstein and Maxwell equations).
Left panels: Density contour plots
show a transition to a toroidal topology as the iterations
proceed. Center panels: ``Potential'' plots similar to Figure 2; note
the increasingly narrow well into which the matter is compressed.
Right panels: ``Force'' plots similar to Figure 4; the tendency of
the gravitational force to join the magnetic force in pushing matter
away from the origin becomes more and more pronounced.}\label{runaway}
\end{figure}

\begin{figure}
\epsscale{0.6}
\plotone{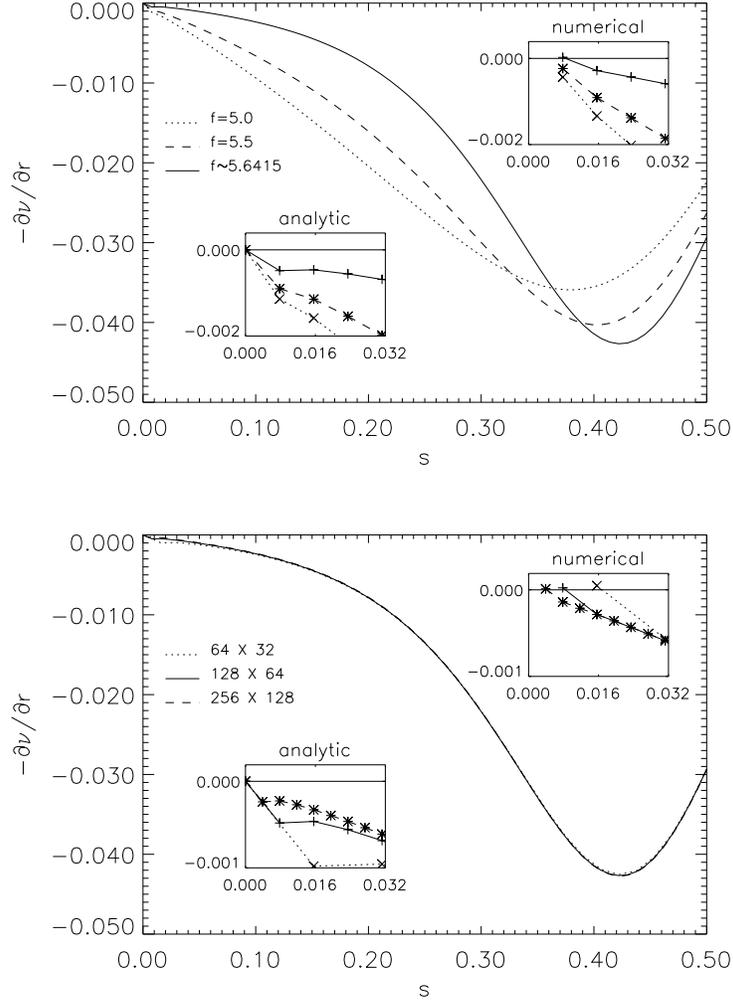}
\caption{The gravitational force in the equatorial plane. 
Upper panel: The gravitational force for increasing values
of $f_0$, at a fixed value of maximum stellar density. The indicated
values of $f_0$ are given in units of $10^{15}
 (R_*\, e_0\, c^2)^{-1}\ {\rm A\ m^{-2} }$, where $R_*$ is the coordinate
radius of the equatorial surface. The insets show
the force near the origin, computed both ``analytically'' and
numerically, as described in the text. Lower panel: The gravitational
force for the maximum value of $f_0$ for which convergence was 
achieved (for a particular value of maximum density), for configurations
computed with three different resolutions in $r$ and $\theta$.}
\label{dnudr}
\end{figure}

\begin{figure}
\epsscale{0.55}
\plotone{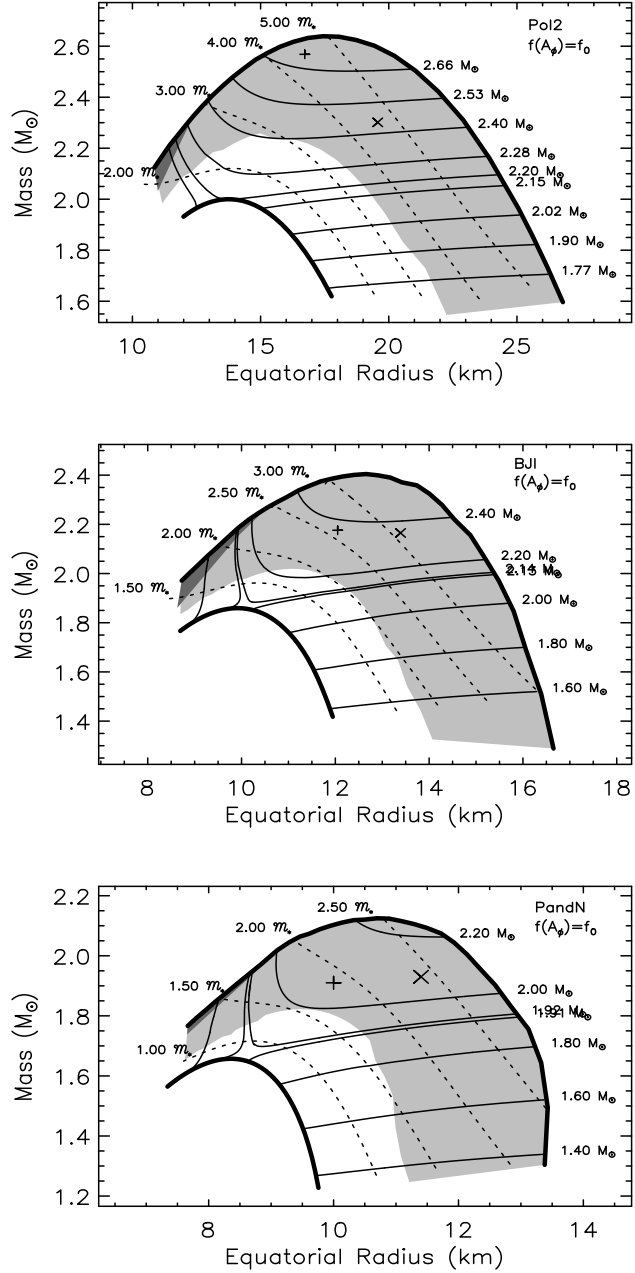}
\caption{Mass-equatorial radius plots showing converged solutions
attainable with a constant current function, for EOSs used by
\cite{bocq95}.  The lower heavy line represents spherical,
non-magnetized, configurations, and the upper heavy line represents
the boundary beyond which solutions appear not to exist (see
\S\ref{sec:himass}).  Lighter solid lines are sequences of constant
baryon mass (in M$_\odot$), while dotted lines are sequences of
constant ${\cal M}$ (in units of ${\cal M_*}=10^{35}$ Gaussian). 
Lighter shaded regions indicate configurations in which the
maximum density is not at the center; the darker shaded regions
indicate gravitationally unbound configurations. ``X''s denote the
maximum mass configuration attainable by uniform rotation, and crosses
indicate the maximum masses for non-rotating, magnetized
configurations reported by \cite{bocq95}.}\label{mr1}
\end{figure}

\begin{figure}
\epsscale{0.6}
\plotone{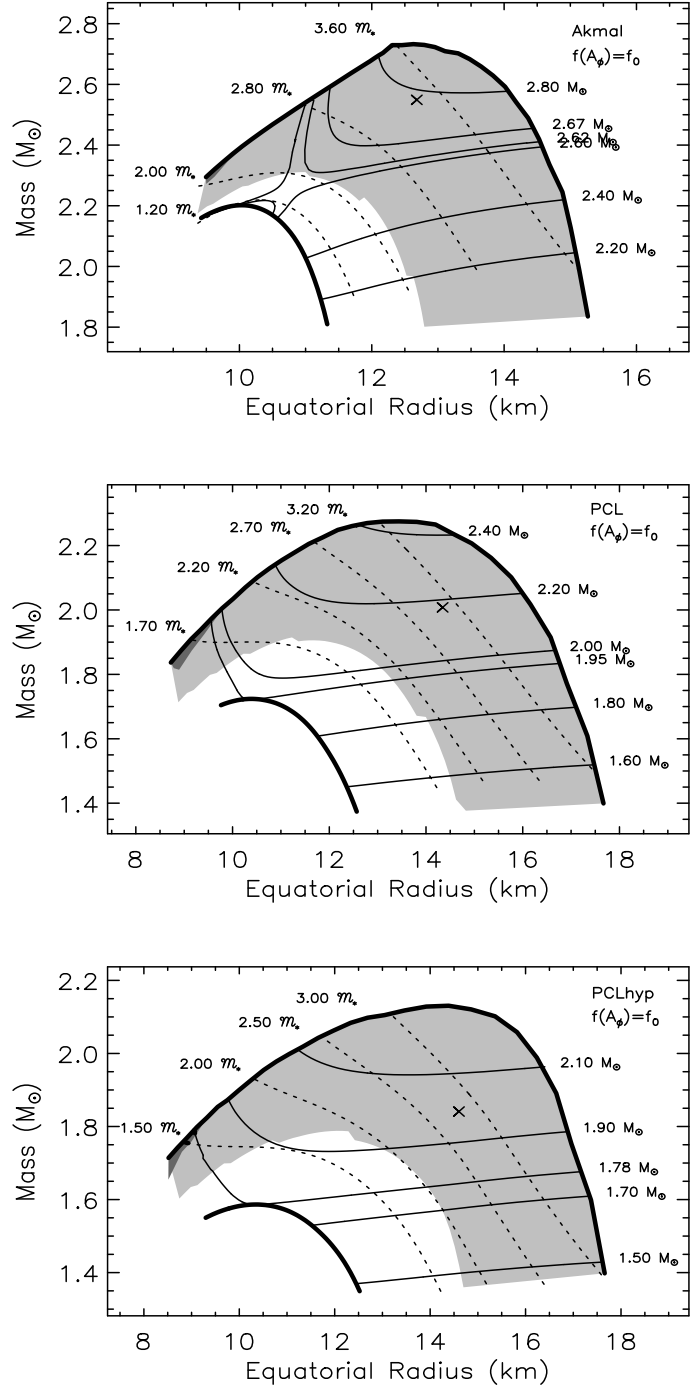}
\caption{Similar to Figure \ref{mr1}, but for three 
more recent EOSs discussed in the text.}\label{mr2}
\end{figure}

The results of these calculations are presented in Table \ref{tbl-2}.
For each EOS, three sets of calculations are presented. The
first set contains results of stars without magnetic fields---the
same data appearing in Table \ref{tbl-1}---as a baseline. The second set
shows results for the maximum mass at a given fixed (relatively low)
value of magnetic moment. The third set for each EOS has two entries 
of our calculations. These two configurations represent the
boundaries of the range of configurations, at fixed magnetic moment, 
whose baryon mass rounds to the value reported by \cite{bocq95} as
the maximum baryon mass among all configurations with constant current
function. The results are satisfactorily concordant. 
For a visual comparison, our Figure \ref{contour} can be checked 
against Figures 5 and 6 of \cite{bocq95}.

We have cited three classes of tests which validate our code. 
First, the quantity
$|1-\lambda|$ is close to zero as required of valid solutions. Second,
our results for quantities characterizing the maximum mass configuration
among uniformly rotating stars agree well with those of two previous
groups. Third, quantities characterizing certain configurations
with magnetic fields reported by \cite{bocq95} show good agreement.

\clearpage

\begin{deluxetable}{lcccccll}
\tabletypesize{\scriptsize} 
\tablecaption{Properties of soft
$\gamma$-ray repeaters (SGRs) and anomalous X-ray pulsars (AXPs) from
the recent literature$^*$.  Question marks indicate uncertain or
unconfirmed values.  
\label{tbl-0}} 
\tablewidth{0pt} 
\tablehead{
\colhead{SGR} & \colhead{P} & \colhead{$\dot P$ \tablenotemark{a}} &
\colhead{$B$ \tablenotemark{b}} & \colhead{$D$ \tablenotemark{c}} &
\colhead{$L_X$ \tablenotemark{d}} & \colhead{Supernova} &
\colhead{Comments} \\ 
&(s)& $(10^{-11}$ s/s) &$(10^{14}$ G) &(kpc) & $(10^{34}$ erg/s) & 
Remnant (SNR) & } 
\startdata 
1806-20 & 7.48 & 2 &
8 & 14 & 20 & G10.0-0.3 & \\ 1900+14 & 5.16 & 6.1 & 8 & 5 & 3 &
G42.8+0.6? & 27-Aug-98 giant flare; \\ & & & & & & & radio point
source \\ 0525-66 & 8 & & 4.5 & 55 & $\sim300$ & N49 in LMC &
05-Mar-79 giant flare \\ 1627-41 & 6.4? & & & 11 & 10 & G337.0-0.1 &
\\ 1801-23 & & & & 10? & & & very recent; only two bursts \\
\tableline AXP & & & & & & & \\ \tableline 4U 0412+61 & 8.69 & 0.23 &
2.8 & 4 & 112 & none & $L_{\rm BB}\sim 40\%\ L_X$.  $\dot P$
constant. \\ 1E 1048.1-5937 & 6.45 & 1.67/3.29/1.67 & 6.5 & 10 & 8 &
none & $L_{\rm BB}\sim 55\%\ L_X$. Three epochs \\ & & & & & & & of
different, but constant, $\dot P$.\\ 1RXS J170849-4009 & 11.00 & 1.9 &
9.2 & 10 & 140 & none & $L_{\rm BB}\sim 55\%\ L_X$. Regular \\ & & & &
& & & spindown, except for a glitch. \\ 1E 1841-045 & 11.77 & 4.1 & 14
& 7 & 35 & Kes 73 & \\ AX J1845.0-0258 & 6.97 & & & 15 & 30 &
G29.6+0.1 & No $\dot P$ to date but young SNR, \\ & & & & & & & hence
large $\dot P$.\\ 1E 2259+586 & 6.97 & 0.06 & 1.3 & 4 & 3.3 & CTB 109
& $L_{\rm BB}\sim 40\%\ L_X$. Bumpy \\ & & & & & & & spin-down.\\
\tableline Magnetar candidates & & & & & & & \\ \tableline RX
J0720.4-3125 & 8.39 & & & 0.1 & 0.0026 & none & Blackbody (BB)
spectrum,\\ & & & & & & & proposed old magnetar \\ RX J0420.0-5022 &
22.7? & & & 4 & 0.4 & none & Needs confirmation. \\ & & & & & & &
Candidate because of large $P$.\\ \enddata 

\tablenotetext{*}{The entries in this table are extracted from
\cite{cline00,corb99,gaen99,gott98,gott99,hab97,hab00,hurl00a,hurl00b,heyl98a,kasp00,kouv98,kouv99,oost98,parm98,paul00,roth94,sugi97,tori98,vasi97a,vasi97b,whit96,wood99a,wood99b}}
\tablenotetext{a}{In most cases there is not enough sampling of $\dot P$
	to allow strong claims of constancy in time, and in many cases
	there is evidence of significant variations.}
\tablenotetext{b}{$B$ is obtained by standard magnetodipolar radiation
	braking, only good to within an order of magnitude.}
\tablenotetext{c}{Distances are poorly determined (with the exception of
	SGR 0525-66 in the LMC).}
\tablenotetext{d}{The X-ray luminosity $L_X$ scales as $D^2$ and is
	estimated from the observed unabsorbed flux using the quoted 
	distance.}
\end{deluxetable}

\begin{deluxetable}{lccccccccc}
\tabletypesize{\scriptsize}
\tablecaption{Maximum mass models at various fixed 
values of magnetic moment. 
\label{tbl-25}}
\tablewidth{0pt}
\tablehead{
\colhead{EOS}  &\colhead{${\cal M}$ } 
	& \colhead{$h_{\rm max}$}   &
\colhead{$e_{\rm max}$ } &
\colhead{$B_c$}  & \colhead{$B_{\rm pole}$} &
	 \colhead{$M$} & 
	\colhead{$M_B$} &
\colhead{$R$}    &
\colhead{$|1-\lambda|$} \\
 &  (10$^{35}$ Gaussian) & & $(1.66\times 10^{14}$ g cm$^{-3}$)& (10$^{16}$ G) 
& (10$^{16}$ G)& 
     (M$_\odot$)	& (M$_\odot$) &  (km) &
}
\startdata
Pol2 &0.00& 0.492 & 10.5 &0.00& 0.00 & 2.00  & 2.19  & 13.8  & 1E-4 \\
     & 2.00 & 0.459 & 9.41 & 150 & 36.9 & 2.12 & 2.28 & 13.9 & 6E-5 \\
     & 3.00 & 0.511 & 11.1& 253 & 122 & 2.37 & 2.38 & 12.8 & 2E-4 \\
     & 4.00 & 0.386 & 7.28& 180 & 92.9 & 2.56 & 2.66 & 15.2 & 3E-4 \\
& {\em 4.99} &{\em 0.291}& {\em 4.94}& {\em 129} &{\em 70.6}&{\em 2.64} &{\em 2.78} &{\em 17.6}  & {\em 3E-4} \\ 
     & 5.00 & 0.290 & 4.92& 129 & 70.4 & 2.64 & 2.78 & 17.6 & 3E-4 \\
& {\em 6.18} &{\em 0.143}& {\em 2.07}& {\em 59.6} &{\em 34.4 }&{\em 2.31} &{\em 2.43} &{\em 22.9}  & {\em 3E-4} \\ 
\tableline
BJI & 0.00& 0.687 & 18.5 &0.00& 0.00 & 1.86 & 2.14 & 9.92   & 5E-4 \\
 & 1.50 & 0.568 & 14.6& 236 & 84.1 & 1.96 & 2.17 & 10.3 & 4E-5 \\
 & 2.00 & 0.706 & 19.2& 419 & 249 & 2.12 & 2.05 & 9.53 & 4E-4 \\
 & 2.50 & 0.586 & 15.1& 335 & 216 & 2.28 & 2.29 & 10.7 & 5E-4 \\
 & 3.00 & 0.476 & 11.9& 269 & 181 & 2.38 & 2.47 & 11.9 & 6E-4 \\
& {\em 3.29} &{\em 0.408}& {\em 10.1}& {\em 232} &{\em 157}&{\em 2.40} &{\em 2.54} &{\em 12.6}  & {\em 1E-3} \\ 
& {\em 3.70} &{\em 0.265}& {\em 6.68}& {\em 156 } &{\em 104}&{\em 2.25} &{\em 2.41} &{\em14.4 }  & {\em 2E-4} \\ 
\tableline
PandN  &0.00& 0.728 & 24.9 &0.00& 0.00 & 1.66 & 1.92 & 8.36  & 5E-5 \\
 & 1.00 & 0.617 & 20.3& 292 & 87.8 & 1.72 & 1.92 & 8.72 & 8E-5 \\
 & 1.50 & 0.724 & 24.7& 504 & 303 & 1.86 & 1.80 & 8.18 & 3E-4 \\
 & 2.00 & 0.590 & 19.3& 386 & 271 & 2.05 & 2.06 & 9.37 & 4E-4 \\
& {\em 2.49} &{\em 0.422 }& {\em 13.7}& {\em 281} &{\em 202}&{\em 2.13} &{\em 2.23} &{\em10.8 }  & {\em 5E-4} \\ 
 & 2.50 & 0.410 & 13.4& 276 & 197 & 2.12 & 2.23 & 10.8 & 5E-4 \\
& {\em 2.82} &{\em 0.231 }&{\em 8.31} & {\em 166} &{\em 119} &{\em 1.90} &{\em2.03 }&{\em 12.6}  & {\em 2E-4} \\ 
\tableline
Akmal & 0.00 & 0.910 & 16.7& 0.00 & 0.00 & 2.20 & 2.67 & 10.0 & 2E-4 \\
 & 1.20 & 0.792 & 14.3& 228 & 59.3 & 2.22 & 2.62 & 10.3 & 7E-5 \\
 & 2.00 & 0.648 & 11.8& 338 & 134 & 2.31 & 2.58 & 10.6 & 3E-4 \\
 & 2.80 & 0.674 & 12.3& 370 & 244 & 2.53 & 2.59 & 11.0 & 2E-4 \\
 & 3.60 & 0.550 & 10.3& 287 & 221 & 2.73 & 2.84 & 12.4 & 3E-4 \\
& {\em 3.71} &{\em 0.514 }&{\em 9.75} & {\em 271} &{\em 209} &{\em 2.73} &{\em 2.88}&{\em12.6 }  & {\em 3E-4} \\ 
& {\em 4.10} &{\em 0.307 }& {\em 6.81}& {\em 179} &{\em 132} &{\em 2.52} &{\em 2.73}&{\em14.3  }  & {\em 1E-3} \\ 
\tableline
PCL & 0.00 & 0.561 & 17.3& 0.00 & 0.00 & 1.72 & 1.95 & 10.4 & 4E-4 \\
 & 1.70 & 0.665 & 22.7& 418 & 218 & 1.91 & 1.86 & 9.14 & 4E-4 \\
 & 2.20 & 0.547 & 16.6& 327 & 186 & 2.09 & 2.12 & 10.4 & 6E-4 \\
 & 2.70 & 0.445 & 12.5& 259 & 155 & 2.21 & 2.31 & 11.7 & 8E-4 \\
 & 3.20 & 0.350 & 9.22& 201 & 124 & 2.26 & 2.41 & 13.1 & 5E-4 \\
& {\em 3.31} &{\em 0.335 }&{\em 8.75} & {\em 192} &{\em 120} &{\em 2.28} &{\em 2.43}&{\em13.4  }  & {\em 4E-4} \\ 
& {\em 3.86} &{\em 0.202 }&{\em 5.12} & {\em 116} &{\em 74.7} &{\em 2.09} &{\em 2.24}&{\em 15.8 }  & {\em 1E-4} \\ 
\tableline
PCLhyp & 0.00 & 0.504 & 17.6& 0.00 & 0.00 & 1.59 & 1.78 & 10.3 & 4E-4 \\
 & 1.50 & 0.620 & 24.3& 416 & 202 & 1.76 & 1.73 & 8.82 & 5E-4 \\
 & 2.00 & 0.488 & 16.7& 308 & 161 & 1.93 & 1.98 & 10.3 & 5E-4 \\
 & 2.50 & 0.387 & 12.1& 235 & 130 & 2.04 & 2.15 & 11.8 & 3E-4 \\
 & 3.00 & 0.310 & 8.97& 182 & 106 & 2.11 & 2.25 & 13.2 & 8E-6 \\
& {\em 3.39} &{\em 0.263 }&{\em 7.29} & {\em 152} &{\em 93.1} &{\em 2.13} &{\em 2.28}&{\em 14.2 }  & {\em 2E-4} \\ 
& {\em 3.81} &{\em 0.186 }&{\em 4.83} & {\em 107} &{\em 68.7} &{\em 2.02} &{\em 2.16}&{\em 16.1 }  & {\em 5E-4} \\ 
\enddata

\tablecomments{The quantity $h_{max}$ is the maximum value of the log-enthalpy
per baryon  defined in equation (\ref{heatfunc}), $e_{max}$ is the
maximum energy density, $B_c$ and $B_{pole}$ are the magnetic field
strengths at the star's center and pole, respectively, $M$ is the
gravitational mass, $M_B$ is the star's baryon mass, $R$ is the
equatorial radius, and $|1-\lambda|$ is a function defined in equation
(\ref{lambda}) describing convergence.  The results are grouped by the
equation of state (EOS), each of which is described in the text.
Entries in plain type correspond to the sequences of constant magnetic
moment ${\cal M}$ shown in Figures \ref{mr1} and \ref{mr2}.  For each
EOS there are two italicized rows which correspond, respectively, to
the configurations of maximum mass and maximum magnetic moment among
all configurations with constant current function.}

\end{deluxetable}

\begin{deluxetable}{llcccccccc}
\tabletypesize{\scriptsize}
\tablecaption{Maximum mass models, nonrotating and rotating. \label{tbl-1}}
\tablewidth{0pt}
\tablehead{
\colhead{EOS} & \colhead{Authors}   & \colhead{$h_c$}   &
\colhead{$e_c$  } &
\colhead{$\Omega$ }  & \colhead{$M $} & 
	\colhead{$M_B$} &
\colhead{$R $}     & \colhead{$J/M^2$}  &
\colhead{$|1-\lambda |$} \\
  & 	& 	&($1.66\times 10^{14}$ g cm$^{-3}$) &(10$^4$ s$^{-1}$)	
&	(M$_\odot$)  &(M$_\odot$)	& (km)	&(G$_N$/c)	&
}
\startdata
Pol2 & CST & \nodata & 10.6 & 0.00 & 1.99 & 2.19  & 13.7 & 0.00 & \nodata \\
 & BBGN & 0.491 & 10.4 & 0.00 & 2.00 & 2.19 & 13.8 & 0.00 & 1E-14 \\
 & CPL & 0.492 & 10.5 & 0.00 & 2.00  & 2.19  & 13.8 & 0.00 & 1E-4 \\
\tableline
 & CST & \nodata  & 8.63 & 0.629 & 2.29  & 2.52 & 19.7 & 0.572 & \nodata  \\
 & BBGN & 0.432 & 8.58 & 0.629 & 2.30 & 2.52  & 19.6 & 0.570 & 6E-6 \\
 & CPL & 0.432 & 8.58 & 0.629 & 2.30  & 2.53  & 19.6 & 0.570 & 1E-4  \\
\tableline
BJI & CST & \nodata & 18.5  & 0.00 & 1.86 & 2.16 & 9.93 & 0.00 & \nodata \\
 & BBGN & 0.699 & 18.6 & 0.00 & 1.86 & 2.13 & 9.91 & 0.00 & 2E-6 \\
 & CPL & 0.687 & 18.5 & 0.00 & 1.86 & 2.14 & 9.92 & 0.00 & 5E-4 \\
\tableline
 & CST & \nodata & 16.0 & 1.06 & 2.17 & 2.49 & 13.4 & 0.629 & \nodata \\
 & BBGN & 0.628 & 16.3 & 1.07 & 2.15 & 2.46 & 13.4 & 0.626 & 9E-5 \\
 & CPL & 0.610 & 15.9 & 1.07 & 2.16 & 2.47 & 13.4 & 0.632 & 4E-5 \\
\tableline
PandN & CST & \nodata & 24.9 & 0.00 & 1.66 & 1.92 & 8.37 & 0.00 & \nodata \\
 & BBGN & 0.733 & 24.4 & 0.00 & 1.66 & 1.93 & 8.55 & 0.00 & 2E-4 \\
 & CPL & 0.728 & 24.9 & 0.00 & 1.66 & 1.92 & 8.36 & 0.00 & 5E-5 \\
\tableline
 & CST & \nodata & 21.4 & 1.32 & 1.95 & 2.24 & 11.2 & 0.665 & \nodata \\
 & BBGN & 0.668 & 21.7 & 1.29 & 1.93 & 2.23 & 11.4 & 0.641 & 7E-5 \\
 & CPL & 0.647 & 21.5 & 1.33 & 1.96 & 2.25 & 11.1 & 0.666 & 1E-5 \\
 \enddata

\tablecomments{$J$ is the total angular momentum.  See the caption to
Table \ref{tbl-25} for an explanation of the other quantities
tabulated.  The subscript $c$ refers to central values.  For each EOS,
results from this work (CPL) are compared to those of \cite{cook94a}
and \cite{cook94b} for the polytropic and tabulated EOSs, respectively
(together labelled CST), and to \cite{bocq95} (BBGN).}

\end{deluxetable}

\begin{deluxetable}{llccccccccc}
\tabletypesize{\scriptsize}
\tablecaption{
Comparison of static models with various magnetic moments. 
\label{tbl-2}}
\tablewidth{0pt} \tablehead{ \colhead{EOS} & \colhead{Authors}
&\colhead{${\cal M}$ } & \colhead{$h_c$} & \colhead{$e_c$ } &
\colhead{$B_c$} & \colhead{$B_{\rm pole}$} & \colhead{$M$} &
\colhead{$M_B$} & \colhead{$R$} & \colhead{$|1-\lambda|$} \\
 & &(10$^{35}$ Gaussian) & &($1.66\times 10^{14}$ g cm$^{-3}$)& (10$^{16}$ G) &
(10$^{16}$ G)& (M$_\odot$) & (M$_\odot$) & (km) & } 
\startdata 
Pol2 & CST&0.00& \nodata & 10.6 &0.00& 0.00 & 1.99 & 2.19 & 13.7 & \nodata \\
 &BBGN &0.00& 0.491 & 10.4 &0.00& 0.00 & 2.00 & 2.19 & 13.8 & 1E-14 \\
 &CPL &0.00& 0.492 & 10.5 &0.00& 0.00 & 2.00 & 2.19 & 13.8 & 1E-4 \\
\tableline
 & BBGN & 0.800 & 0.483 & 10.2 & 59.8 & 9.8 & 2.01 & 2.21 &13.8 & 1E-6 \\
 & CPL & 0.800 & 0.484 & 10.2 & 60.4 & 10.0 & 2.02 &2.21 & 13.8 & 2E-4 \\
\tableline
 & BBGN & 4.49 & 0.225 & 3.55 & 142 &72.3 & 2.57 & 2.71 & 16.8 & 1E-3 \\
 & CPL & 4.49 & 0.224 & 3.53 & 140& 70.7 & 2.56 & 2.71 & 16.8 & 2E-4 \\
 & CPL & 4.49 & 0.224 & 3.53 &142 & 72.6 & 2.57 & 2.71 & 16.8 & 3E-4 \\
\tableline
 BJI & CST &0.00&\nodata & 18.5 &0.00& 0.00 & 1.86 & 2.16 & 9.93 & \nodata \\
 & BBGN&0.00& 0.699 & 18.6 &0.00& 0.00 & 1.86 & 2.13 & 9.91 & 2E-6 \\
 & CPL&0.00& 0.687 & 18.5 &0.00& 0.00 & 1.86 & 2.14 & 9.92 & 5E-4 \\
\tableline
 & BBGN & 0.30 & 0.692 & 18.4 & 61.5 & 11.0 & 1.86 & 2.13 & 9.94 & 2E-6 \\
 & CPL & 0.30 & 0.680 & 18.2 & 60.9 & 11.1 & 1.86 & 2.14 & 9.95 & 2E-4 \\
\tableline
 & BBGN & 2.63 & 0.300 & 7.47 & 233 & 121 & 2.18 & 2.34 & 12.0 & 1E-4 \\
 & CPL & 2.63 & 0.287 & 7.16 & 215 & 110 & 2.15 & 2.34 & 12.3 & 2E-4 \\
 & CPL & 2.63 & 0.292 & 7.27 & 222 & 115 & 2.17 & 2.34 & 12.2 & 2E-4 \\ 
\tableline
PandN & CST & 0.00&\nodata & 24.9 &0.00& 0.00 & 1.66 & 1.92 & 8.37 & \nodata \\
 & BBGN &0.00& 0.733 & 24.4 &0.00& 0.00 & 1.66 & 1.93 & 8.55 & 1E-4 \\
 & CPL &0.00& 0.728 & 24.9 &0.00& 0.00 & 1.66 & 1.92 & 8.36  & 5E-5 \\
\tableline
 & BBGN & 0.20 & 0.727 & 24.1 & 64.7 & 11.9 & 1.66 & 1.93 & 8.57 & 1E-4 \\
 & CPL & 0.20 & 0.722 & 24.6 & 66.6 & 12.9 & 1.66 & 1.92 & 8.38 & 4E-5 \\
\tableline
 & BBGN & 1.86 & 0.350 & 11.2 & 303 & 154 & 1.91 & 2.06 & 10.0 & 3E-4 \\
 & CPL & 1.86 & 0.328 & 11.0 & 292 & 153 & 1.90 & 2.06 & 10.0 & 3E-4 \\
 & CPL & 1.86 & 0.336 & 11.2 & 361 & 209 & 1.96 & 2.06 & 9.47 & 3E-4
 \enddata

\tablecomments{See the notes to Tables \ref{tbl-25} and \ref{tbl-1}
for explanations of symbols and abbreviations.}

\end{deluxetable}

\clearpage

\begin{thebibliography}{}
\bibitem[Akmal et al.(1998)]{akmal98}Akmal, A.,
	Pandharipande, V. R., \& Ravenhall, D. G. 1998, 
	Phys. Rev. C 58, 1804 
\bibitem[Alv\'en(1950)]{alfv50}Alfv\'en, H. 1950, Cosmical 
	Electrodynamics (Oxford: Clarendon)
\bibitem[Bardeen \& Wagoner(1971)]{bard71}Bardeen, J. M.  
	\& Wagoner, R. V. 1971,
	ApJ, 167, 359
\bibitem[Baumgarte, Shapiro, \& Shibata(2000)]{baum00}Baumgarte,
	T. W., Shapiro, S. L., \& Masaru, S. 2000, \apj, 528, L29
\bibitem[Baym, Bethe, \& Pethick(1971)]{baym71a}Baym, G., Bethe, H. A.,
	\& Pethick, C. 1971, Nucl. Phys. A, 175, 225
\bibitem[Baym, Pethick, \& Sutherland(1971)]{baym71b}Baym, G., Pethick, C.,
	\& Sutherland, P. 1971, \apj, 170, 299
\bibitem[Bethe \& Johnson(1974)]{beth74}Bethe, H. A. \& Johnson, M. B. 1974,
	Nucl. Phys. A, 230, 1
\bibitem[Bocquet et al.(1995)]{bocq95}Bocquet, M., 
	Bonazzola, S., Gourgoulhon, E.,
	\& Novak, J. 1995, A\&A, 301, 757 (BBG)
\bibitem[B\"ohm-Vitense(1989)]{bohm89}B\"ohm-Vitense, E. 1989, 
	Introduction to Stellar
	Astrophysics, Vol. 1 (Cambridge: Cambridge UP), Ch. 14
\bibitem[Bonazzola, Gourgoulhon, \& Marck(1998)]{bona98} Bonazzola, S., 
	Gourgoulhon, E., \&
	Marck, J. A. 1998, Phys. Rev. D, 58, 104020
\bibitem[Bonazzola et al.(1993)]{bona93} Bonazzola, S., 
	Gourgoulhon, E., Salgado, M., \&
	Marck, J. A. 1993, A\&A, 278, 421
\bibitem[Bonazzola \& Maschio(1971)]{bona71} Bonazzola, S. 
	\& Maschio, G. 1971, in
	The Crab Nebula, ed. R. D. Davies \& F. G. Smith
	(Dordrecht: Reidel), 346
\bibitem[Bonazzola \& Schneider(1974)]{bona74} Bonazzola, S. 
	\& Schneider, J. 1974, ApJ,
	191, 273
\bibitem[Broderick, Prakash, \& Lattimer(2000)]{bpl00} Broderick, A., 
	Prakash, M., \& Lattimer, J. M. 2000, ApJ, 537, 351
\bibitem[Cardall, Broderick, Prakash, \& Lattimer(2000)]{card00}
	Cardall, C. Y., Broderick, A., Prakash, M., \& Lattimer, J. M. 2000,
	to be published 
\bibitem[Carroll \& Ostlie(1996)]{carr96}Carroll, B. W. \& Ostlie, D. A. 1996,
	 An Introduction
	to Modern Astrophysics (Reading: Addison-Wesley), \S 15.6
\bibitem[Carter(1973)]{cart73}Carter, B. 1973, in Black Holes, ed. 
	C. DeWitt \& B. S. DeWitt (New York: Gordon and Breach), 125
\bibitem[Chakrabarty(1996)]{cha96} Chakrabarty, S. 1996, 
	Phys. Rev. D, 54, 1306
\bibitem[Chakrabarty, Bandyopadhyay, \& Pal(1997)]{cha97} 
	Chakrabarty, S., Bandyopadhyay, D., \& Pal, S. 1997, 
	Phys. Rev. Lett., 78, 2898
\bibitem[Chandrasekhar \& Fermi(1953)]{chan53}Chandrasekhar, S. \&
	Fermi, E. 1953, \apj, 118, 116
\bibitem[Chatterjee, Hernquist, \& Narayan(2000)]{chat00}Chatterjee, P.,
	Hernquist, L., \& Narayan, R. 2000, \apj, 534, 373

\bibitem[Cline, et al.(2000)]{cline00}
Cline, T., Frederiks, D. D., Golenetskii, S., Hurley, K., 
Kouveliotou, C., Mazets, E., van Paradijs, J. 2000, \apj, 531, 407
\bibitem[Cook, Shapiro, \& Teukolsky(1992)]{cook92} Cook, G. B., 
	Shapiro, S. L., \& Teukolsky, S. A.
	1992, ApJ, 398, 203
\bibitem[Cook, Shapiro, \& Teukolsky(1994a)]{cook94a} Cook, G. B., 
	Shapiro, S. L., \& Teukolsky, S. A.
	1994a, ApJ, 422, 227
\bibitem[Cook, Shapiro, \& Teukolsky(1994b)]{cook94b} Cook, G. B., 
	Shapiro, S. L., \& Teukolsky, S. A.
	1994b, ApJ, 424, 823


\bibitem[Corbel, Chapuis, Dame, \& Durouchoux (1999)]{corb99}
Corbel, S., Chapuis, C., Dame, T. M., \& Durouchoux, P. 1999, \apj, 
526, L29 

\bibitem[Cowling(1934)]{cowl34}Cowling, T. G. 1934, \mnras, 94, 39
\bibitem[Duncan \& Thompson(1992)]{dunc92} Duncan, R. C. \& 
	Thompson, C. 1992, ApJ, 392, L9

\bibitem[Friedman, Ipser, \& Parker(1986)]{Frie86}
Friedman, J. L, Ipser, J. R., \& Parker, L., \apj, 304, 115

\bibitem[Friedman, Ipser, \& Sorkin(1988)]{frie88}Friedman, J. L.,
	Ipser, J. R., \& Sorkin, R. D. 1988, \apj, 325, 722



\bibitem[Gaensler, Gotthelf, \& Vasisht(1999)]{gaen99}
Gaensler, B. M., Gotthelf, E. V. \& Vasisht, G. 1999, \apj, 526, L37 


\bibitem[Goldreich \& Reisenegger(1992)]{gold92}Goldreich, P., 
	\& Reisenegger, A.
	1992, \apj, 395, 250


\bibitem[Gotthelf \& Vasisht(1998)]{gott98}
Gotthelf, E. V. \& Vasisht, G. 1998, New Astr. 3, 293 

\bibitem[Gotthelf, Vasisht, \& Dotani(1999)]{gott99}
Gotthelf, E. V., Vasisht, G. \& Dotani, T. 1999, ApJ 522, L49 

\bibitem[Haberl, Motch, Buckley,  Zickraf, \& Pietsch(1997)]{hab97}
Haberl, F., Motch, C., Buckley, D. A. H., Zickraf, F.-J., 
\& Pietsch, W. 1997, A\&A, 326, 662 

\bibitem[Haberl, Pietsch, \& Motch(2000)]{hab00}
Haberl, F., Pietsch, W. \& Motch, C. 2000, A\&A 351, L53 

\bibitem[Harding et al.(1999)]{hard99}Harding, A. K., Contopoulos, I.
	\& Kazanas, D. 1999, \apj, 525, L125

\bibitem[Heyl \& Hernquist(1998)]{heyl98a}
Heyl, J. S. \& Hernquist, L. 1998, MNRAS, 297, L69
 
\bibitem[Heyl \& Kulkarni(1998)]{heyl98b}Heyl, J. S. \& Kulkarni,
	S. R. 1998, \apj, 506, L61

\bibitem[Hurley(2000)]{hurl00a}
Hurley, K.: 2000, 
in R.~M. Kippen, R.~S. Mallozzi, \& G.~J. Fishman (eds.),  {\it
  Gamma-Ray Bursts}, in press; astro--ph/9912061 

\bibitem[Hurley et al.(2000)]{hurl00b}
Hurley, K., Strohmayer, T., Li, P.,  Kouveliotou, C.,
Woods, P., van Paradijs, J., Murakami, T., Hartmann, D., 
Smith, I., Ando, M., Yoshida, A., Sugizaki, M.
 2000, ApJ, 528, L21 

\bibitem[Imamura, Durisen, \& Friedman(1985)]{Imam85}
Imamura, J., Durisen, R. H., \& Friedman, J. L. 1985, 
\apj, 294, 474 

\bibitem[Ipser, \& Lindblom(1989)]{Ipse89}
Ipser, R., \& Lindblom, L. 1989, Phys. Rev. Lett. 62, 2777
 

\bibitem[Kaspi, Lackey, \& Chakrabarty(2000)]{kasp00}
Kaspi, V. M., Lackey, J. R. \& Chakrabarty, D. 2000, \apj, 537, L 31 


\bibitem[Komatsu, Eriguchi, \& Hachisu(1989)]{koma89}Komatsu, H.,
	Eriguchi, Y., \& Hachisu, I. 1989, \mnras, 237, 355
\bibitem[Kouveliotou et al.(1994)]{kouv94}Kouveliotou, C., 
Bishman, G. J.,
B Meegan, C. A., Paciesas, W. S.,
B van Paradijs, J.; Norris, J. P.,
B Preece, R. D., Briggs, M. S.,
B Horack, J. M., Pendleton, G. H.,
B Green, D. A. 1994, Nature, 368, 125

\bibitem[Kouveliotou et al.(1998)]{kouv98}Kouveliotou, C. 
Dieter, S., Strohmayer, T., van Paradijs, J., Fishman, G. J., 
Meegan, C. A., Hurley, K. 	
1998, Nature, 393, 235

\bibitem[Kouveliotou et al.(1999)]{kouv99}Kouveliotou, C. 
Strohmayer, T., Hurley, K.,
van Paradijs, J., Finger, M. H.,
Dieters, S., Woods, P.,
Thompson, C., Duncan, R. C.
1999, \apj, 510, L115

\bibitem[Lindblom, \& Detweiler(1977)]{Lind77}
Lindblom, L., \& Detweiler, S. L. 1977, \apj, 211, 565

\bibitem[Malone, Johnson, \& Bethe(1975)]{malo75}Malone, R. C., 
	Johnson, M. B., \& Bethe, H. A. 1975, \apj, 199, 741 
\bibitem[Marsden et al.(1999a)]{mars99a}Marsden, D., Lingenfelter, R. E., 
	Rothschild, R. E., \& Higdon, J. C. 1999a, preprint
	(astro-ph/9912207)
\bibitem[Marsden, Rothschild, \& Lingenfelter(1999b)]{mars99b}Marsden,
	D., Rothschild, R. E., \& Lingenfelter, R. E. 1999b,
	ApJ, 520, L107


\bibitem[Oosterbroek, Parmar, Mereghetti,  \& Israel(1998)]{oost98}
Oosterbroek, T., Parmar, A. N., Mereghetti, S. \& Israel, G. L. 1998, 
A\&A 334, 925 

\bibitem[Paczy\'nski(1992)]{pacz92} Paczy\'nski, B. 1992, 
	Acta Astron., 42, 145
\bibitem[Pandharipande(1971)]{pand71}Pandharipande, V. R. 1971,
	Nucl. Phys. A, 174, 641
\bibitem[Parmar et al. (1998)]{parm98}
Parmar, A. N., 
Oosterbroek, T.,
Favata, F., Pightling, S.,
Coe, M. J., Mereghetti, S.,
Israel, G. L. 1998, A\&A 330, 175 

\bibitem[Paul, Kawasaki, Dotani, \& Nagase(2000)]{paul00}
Paul, B., Kawasaki, M., Dotani, T., \& Nagase, F. 2000, 
\apj, 537, 319 

\bibitem[Prakash et al.(1995)]{pcl95}Prakash, M., 
	Cooke, J. R., \& Lattimer, J. M. 1995, Phys. Rev. D 52, 661

\bibitem[Rothschild, Kulkarni, \& Lingenfelter(1994)]{roth94}
Rothschild, R. E., Kulkarni, S. R., \& Lingenfelter, R. E. 1994, 
Nature, 368 


\bibitem[Salgado et al.(1994)]{salg94}Salgado, M., Bonazzola, S., 
	Gourgoulhon, E., \& Haensel, P. 1994, A\&A, 291, 155 

\bibitem[Sawyer(1989)]{Sawy89}
Sawyer, R. F. 1989, Phys. Rev. D 39, 3804 

\bibitem[Shapiro \& Teukolsky(1983)]{shap83}Shapiro, S. L. \& 
	Teukolsky, S. A. 1983, Black Holes, White
	Dwarfs, and Neutron Stars (New York: John Wiley \& Sons), \S 10.5
\bibitem[Shapiro, Teukolsky, \& Winicour(1995)]{shap95}Shapiro, S. L., 
	Teukolsky, S. A., \& Winicour, J. 1995, Phys. Rev. D. 52, 6982
\bibitem[Sorkin(1982)]{sork82}Sorkin, R. D. 1982, \apj, 257, 847
\bibitem[Spruit(1999a)]{spru99a}Spruit, H. C. 1999a, A\&A, 341, L1
\bibitem[Spruit(1999b)]{spru99b}Spruit, H. C. 1999b, A\&A, 349, 189


\bibitem[Sugizaki, et al.(1997)]{sugi97}
Sugizaki, M., 
Nagase, F., 
Torii, K., 
Kinugasa, K., 
Asanuma, T., 
Matsuzaki, K., Koyama, K., 
Yamauchi, S. 1997, Publ. Astron. Soc. Japan., 49, L25 

\bibitem[Taylor, Manchester \& Lyne(1993)]{tayl93}Taylor, J. H., Manchester, 
	R. N., \& Lyne, A. G. 1993, \apjs,
	88, 529
\bibitem[Thompson \& Duncan(1995)]{thom95} Thompson, C. \& Duncan, 
	R. C. 1995, MNRAS, 275, 255
\bibitem[Thompson \& Duncan(1996)]{thom96} Thompson, C. \& Duncan, 
	R. C. 1996, ApJ, 473, 322

\bibitem[Torii et al.(1998)]{tori98}
Torii., K., 
Kinugasa, K., 
Katayama, K., 
Tsunemi, H., 
Yamauchi, S. 1998, ApJ 503, 843 


\bibitem[Usov(1992)]{usov92}Usov, V. V. 1992, Nature, 357, 472
\bibitem[Vasisht \& Gotthelf(1997)]{vasi97a}Vasisht, G. 
	\& Gotthelf, E. V. 1997, \apj, 486, L129


\bibitem[Vasisht, Kulkarni, Anderson, \& Hamilton(1997)]{vasi97b}
Vasisht, G., Kulkarni, S. R., Anderson, S. B., \& Hamilton, T. T. 
1997, \apj, 476, L43 

\bibitem[White et al.(1996)]{whit96}
White, N. E. 
Angelini, L.,
Ebisawa, K., Tanaka, Y.,
Ghosh, P. 1996, \apj, 463, L83 

\bibitem[Woods et al.(1999a)]{wood99a}
Woods, P. M., 
Kouveliotou, C., 
van Paradijs, J., 
Finger, M. H.,
Thompson, C. 
 1999, ApJ, 518, L103 
\bibitem[Woods et al.(1999b)]{wood99b}
Woods, P. M., 
Kouveliotou, C., 
van Paradijs, J., 
Finger, M. H.,
Thompson, C., 
Duncan, R. C.,
Hurley, K.,
Strohmayer, T.,
Swank, J., 
Murakami, T. 
 1999, ApJ, 524, L55
\bibitem[Yuan \& Zhang(1999)]{yz99} Yuan, Y. F., \& Zhang, J. L. 1999, ApJ, 
	525, 920
\bibitem[Zimanyi \& Moszkowski(1990)]{zm90}Zimanyi, J. \& Moszkowski, S. A. 
	1990, Phys. Rev. C 42, 416
\end{thebibliography}
\end{document}